\documentclass[letterpaper,twocolumn,10pt]{article}
\usepackage{usenix}

\usepackage{tikz}
\usepackage{amsmath}
\usepackage[section]{algorithm}
\usepackage{algorithmic}
\usepackage{enumitem}
\usepackage{filecontents}
\usepackage{epsfig,subfigure,wrapfig}
\usepackage{tabularx}
\usepackage{booktabs}       % professional-quality tables
\usepackage{multirow}

\newtheorem{definition}{Definition}[section]

\begin{document}
%-------------------------------------------------------------------------------

%don't want date printed
\date{}

% \title{LayerChunk: Efficient LLM Re-Prefill through Semantic Chunking and Reuse-aware Prefetching}
\title{ContiguousKV: Accelerating LLM Prefill with Granularity-Aligned KV Cache Management}

%for single author (just remove % characters)
\author{
{Jing Zou~\thanks{These authors contribute equally. \dag corresponding author.}}\\
UESTC
\and
{Shangyu $Wu^{*}$}\\
MBZUAI
% copy the following lines to add more authors
\and
{Hancong Duan}\\
UESTC
\and
{Qiao $Li^{\dag}$}\\
MBZUAI
\and
{Chun Jason Xue}\\
MBZUAI
} % end author

\maketitle

%-------------------------------------------------------------------------------
\begin{abstract}
Efficiently serving Large Language Models (LLMs) with persistent Prefix Key-Value (KV) Cache is critical for applications like conversational search and multi-turn dialogue. Serving a request requires loading the pre-computed prefix KV cache and generating the first token, defined as the Re-Prefill Phase. Offloading this shared prefix cache to secondary storage is essential for memory scalability.  Re-Prefill with offloading suffers from severe I/O bottlenecks in two aspects. First, semantic-aware KV cache pruning algorithms select important tokens in fine granularity, while systems manage I/O in coarse, fixed-size blocks, causing severe read amplification. Second, the sequential dependency between identifying important tokens and loading KV cache creates idle I/O and compute bubbles, under-utilizing system resources.

This paper proposes \textit{ContiguousKV}, a high-performance prefix KV cache offloading system that bridges algorithmic semantics with I/O efficiency to accelerate the Re-Prefill phase. We first introduce \textit{ContiguousChunk}, a unified data management granularity that aligns KV cache pruning with I/O operations. All the mechanisms critical for I/O performance are performed at the granularity of ContiguousChunk, thereby eliminating read amplification. By exploiting the high similarity in important ContiguousChunk indices across layers, we propose intra- and inter-period asynchronous prefetching to break the sequential dependency between I/O and compute, effectively eliminating idle bubbles. Finally, we propose attention-guided cache management to retain semantically critical prefix data in memory. Evaluations on Qwen2.5 series models show that ContiguousKV achieves a 3.85x speedup in the Re-Prefill phase over the state-of-the-art offloading system IMPRESS, while maintaining high output quality.
\end{abstract}

\section{Introduction}
A growing class of Large Language Model (LLM) serving workloads, including conversational assistants~\cite{hia, 24arix-stockagent}, retrieval-augmented generation (RAG)~\cite{24arXiv-rag-survey, 24nips-uda}, and interactive document analysis~\cite{25acl-longbench-v2}, is characterized by shared context. 
Multiple user requests (e.g., follow-up questions about a retrieved document) share a large, identical prefix (the document text). 
To avoid redundant computation, systems can persistently store the pre-computed Key-Value (KV) cache of this shared prefix and reuse it across requests~\cite{25fast-impress, 25eurosys-cacheblend, 24atc-as}.
This reuse introduces a pivotal but costly stage: the Re-Prefill Phase.
The Re-Prefill Phase processes a new request by loading the pre-computed KV Cache of a shared prefix from a slower storage tier (e.g., CPU memory or SSD), computing the KV cache of the new, non-shared suffix tokens, and performing attention over the pre-computed KV cache to generate the first token.
% Offloading the large prefix KV cache to secondary storage is necessary for memory scalability, but makes Re-Prefill I/O-bound.

To improve the performance in the Re-Prefill Phase, existing state-of-the-art offloading-based inference systems~\cite{25fast-impress, 24atc-as} identify the critical tokens that dominate model performance and load their pre-computed KV cache from SSDs to perform the inference.
However, a fundamental granularity mismatch exists between algorithmic decisions and system operations.
KV cache selection algorithms operate on semantic units—individual tokens or small, contiguous token groups (chunks) that preserve semantics.
In contrast, offloading systems like IMPRESS~\cite{25fast-impress} manage data in coarse, fixed-size blocks (e.g., 64-token chunks, 1.8MB) to amortize I/O overhead.
This mismatch forces systems to load entire multi-megabyte blocks to access only a few kilobytes of relevant, pruned KV cache, leading to severe read amplification and wasted I/O bandwidth. 
The system’s operational granularity is blind to algorithmic semantics.
In addition, the Re-Prefill computation is inherently serialized per layer.
It must load prefix keys before computation, causing a strict sequential dependency between pruning decisions and KV cache loading.
This leads to large idle ``bubbles'' where either the GPU stalls for I/O or the I/O subsystem idles, indicating resource underutilization with limited Re-Prefill performance.
Furthermore, managing the limited, faster memory as a cache for the offloaded prefix with traditional policies overlooks the semantic value of data, resulting in poor hit rates for the tokens most relevant to ongoing requests.

We argue that accelerating the Re-Prefill phase requires a co-design that: aligns the system’s I/O granularity to the algorithm’s KV cache pruning granularity to eliminate I/O waste, and orchestrates asynchronous, predictive I/O to break the sequential dependency and reclaim idle bubbles.
We present \textbf{ContiguousKV}, a system architected specifically to optimize the Re-Prefill phase. 
ContiguousKV is built around a core abstraction, \textit{ContiguousChunk}, which serves as the unified granularity for importance scoring, storage, transfer, and caching. 
This alignment ensures that every I/O operation fetches precisely the semantically cohesive unit needed by the algorithm, eliminating read amplification.
Building on this, we observe that the set of important token indices in ContiguousChunk exhibits strong similarity across transformer layers. 
We exploit this by designing an intra-period and inter-period asynchronous prefetching engine. 
This engine speculatively fetches KV cache for upcoming layers while the current layer computes, breaking the sequential dependency and transforming the Re-Prefill phase into a pipelined, high-throughput process.
To optimize the use of faster memory tiers, we propose a cache policy that dynamically prioritizes ContiguousChunks with high attention scores, a direct measure of semantic importance for the current request context. 
This ensures the most valuable prefix data remains readily accessible, further reducing the depth and frequency of slow I/O.

We conducted comprehensive experiments on Qwen2.5 models with a large, reusable prefix KV cache.
Experiments demonstrate that ContiguousKV achieves a 3.85x speedup in the Re-Prefill phase compared to state-of-the-art offloading-based inference systems, while maintaining high accuracy.

This paper makes the following contributions:
\begin{itemize}[topsep=0pt,itemsep=-1ex,partopsep=1ex,parsep=1ex]
    \item We introduce the ContiguousChunk as a unified granularity, aligning system I/O to algorithmic pruning to eliminate read amplification during Re-Prefill.
    \item We design a two-level asynchronous prefetching mechanism that exploits cross-layer similarity to pipeline I/Os with model computations, thus eliminating idle bubbles.
    \item We develop an attention-guided cache policy that utilizes semantic importance to enhance cache efficiency.
\end{itemize}

The rest of the paper is organized as follows.
\S \ref{sec:basics} provides background and \S \ref{sec:motivation} presents a detailed motivational analysis of the Re-Prefill bottleneck.
\S \ref{sec:design} introduces the proposed ContiguousKV design.
The experiments are presented in \S \ref{sec:evaluation}.
\S \ref{sec:related} shows the related work and \S \ref{sec:conclusion} concludes this work.
\section{Background}
\label{sec:basics}
\subsection{Modern LLMs Basics}

\begin{figure}[t]
\centering
  \includegraphics[width=1\linewidth]{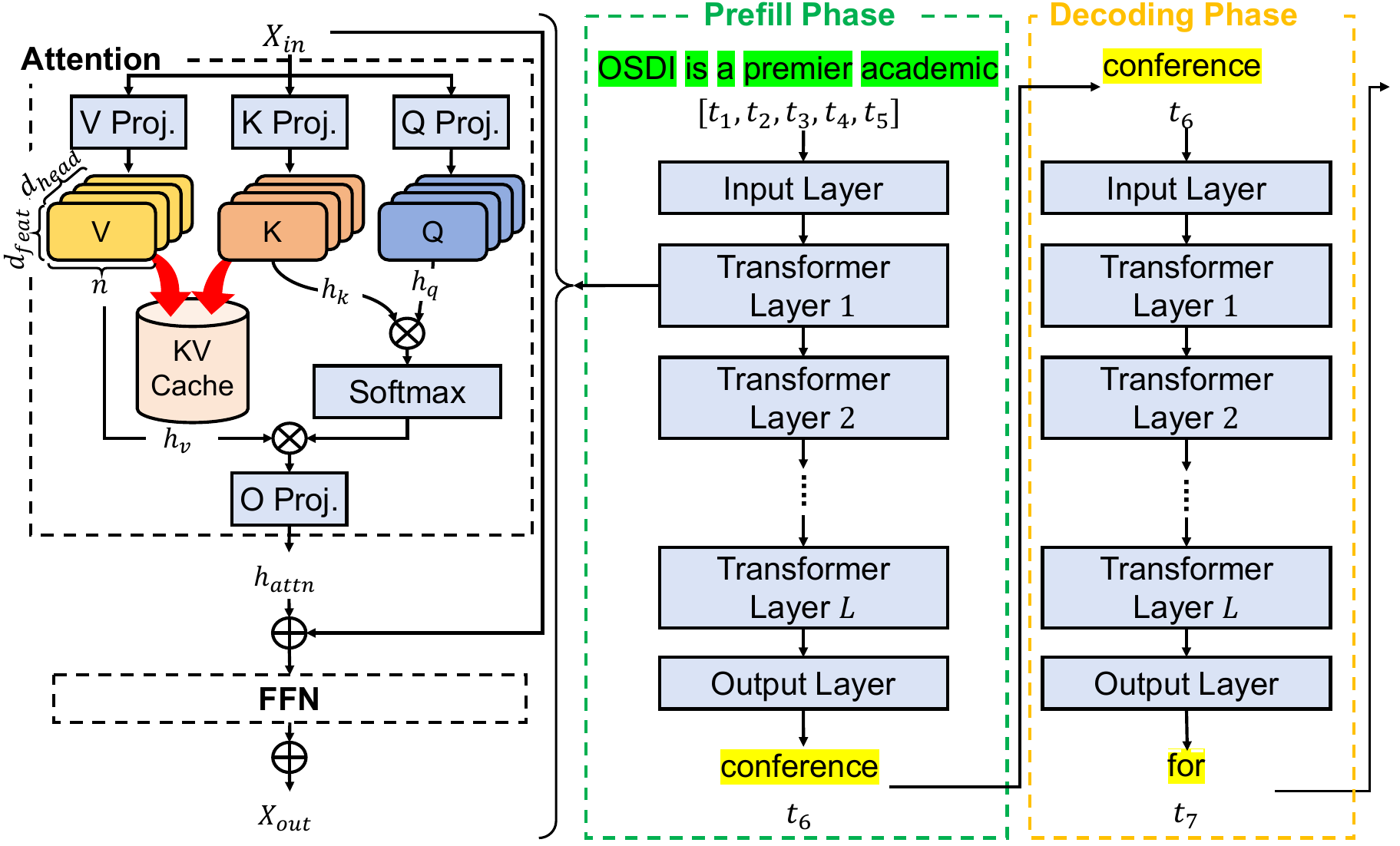}
  \caption{The standard architecture of modern LLMs and the inference phases (i.e., prefill phase and decoding phase).}
  \label{fig:llm-arch}
\end{figure}

\textbf{LLM Architecture and Computation Flows.}
As shown in Figure~\ref{fig:llm-arch}, a typical large language model~\cite{24arxiv-llama3, 22arxiv-opt, 24arxiv-qwen2.5} comprises an input layer, an output layer, and stacks of Transformer layers.
Each Transformer layer primarily consists of a self-attention module and a feed-forward network (FFN) module.
The self-attention module projects each token's hidden state into Query (Q), Key (K), and Value (V) tensors, allowing the model to assess the importance of each token.
The FFN, typically a multi-layer perceptron, then non-linearly transforms each token's representation independently. 
This architecture enables LLMs to capture complex, contextual relationships within the inputs.

Formally, given an input sequence $T=[t_1, t_2, \ldots, t_n]$ of $n$ tokens, the model computes the first output token $t_{n+1}$ through a forward pass across all $L$ layers.
First, the input tokens $T$ are transformed into an input tensor $X_{in}\in\mathbf{R}^{n\times d_{feat}}$, where $d_{feat}$ is the feature dimension of LLM.
Then, $X_{in}$ is passed to the Transformer layers and the output layer to obtain the final output $X_{out}$.
For layer $L_l$, assume that the input tensor is $X^l_{in}$, then it will be multiplied by three weight matrices (i.e., $W^l_q$, $W^l_k$, and $W^l_v$) to generate the Q, K, and V tensors (i.e., $h^l_q$, $h^l_k$, and $h^l_v$) in the attention modules.
These tensors are of shape $n \times d_{head} \times d_{feat}$, where the $d_{head}$ dimension is used to capture different syntactic dependencies, semantic roles, entity relationships, and so on~\cite{25acl-head, 25iclr-head, 23emnlp-f-head}.
Then, $h^l_q$, $h^l_k$, and $h^l_v$ are used to compute the attention output, $h^l_{attn}=\left(\text{softmax}(h^l_q \cdot h^l_k)\cdot h^l_v\right) \cdot W^l_o$, where $W^l_o$ is the weight matrix of output projection.
After the residual connection, the attention output serves as the input of the FFN module.
The FFN module primarily utilizes two linear projections with an activation function in between to derive the outputs of a Transformer layer.
We omit the details of the FFN module as it is not the focus of this paper.

\noindent\textbf{Inference Phases.}
The inference process of an LLM can be decomposed into two distinct phases, i.e., the prefill phase and the decoding phase, as illustrated in Figure~\ref{fig:llm-arch}.
(1) \textbf{Prefill Phase}: the inference begins with this phase, which processes the entire input sequence as the prompt input of the LLM.
For each layer, the generated key and value tensors are stored as a Key-Value (KV) Cache, providing context information for the decoding phase.
% This phase is typically compute-bound and highly parallelizable, as it involves large matrix multiplications resulting from batching multiple inputs.
The output of this phase is the first generated token, and the primary performance metric is the Time-To-First-Token (TTFT).
(2) \textbf{Decoding Phase}: Based on the cached KVs from the prefill phase, the model starts to generate new tokens auto-regressively.
At each step, the model concatenates the cached KVs with newly generated KVs and then performs attention to produce the next token.
% This phase is memory-bound due to the growing KV cache, and is usually measured by throughput (i.e., tokens per second).
Although several works~\cite{24atc-as, 25fast-impress, 25sigmod-pqcache, 25acl-f-a2ats, 25acl-clusterattn} have addressed the memory-bound challenge by offloading the KV cache to main memory or disks, the new I/O-bound challenge arises.

\begin{figure}[t]
\centering
  \includegraphics[width=1\linewidth]{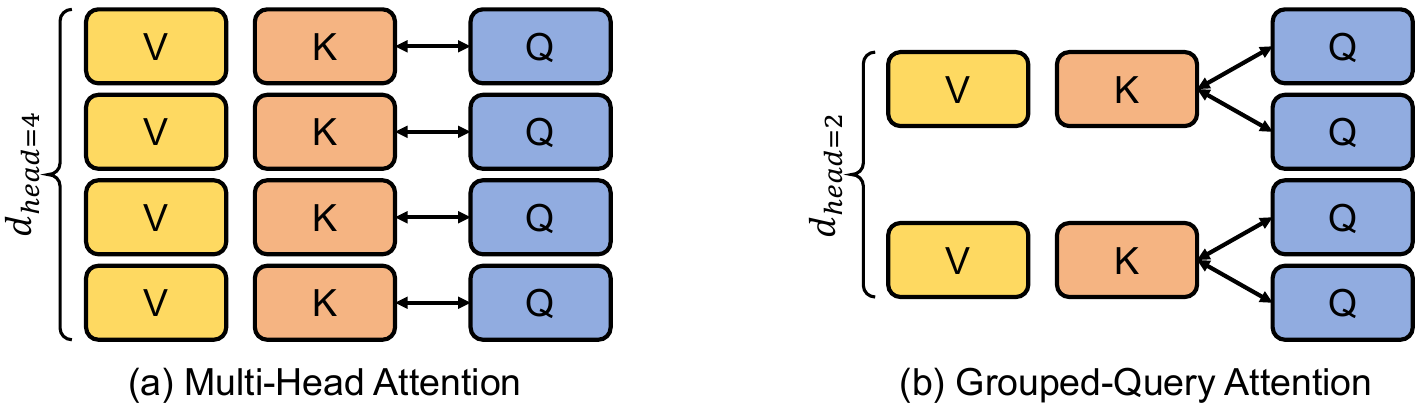}
  \caption{Comparison between multi-head attention and grouped-query attention.}
  \label{fig:gqa}
\end{figure}

\noindent\textbf{Emerging LLM Architectures: Grouped-Query Attention (GQA).}
As shown in Figure~\ref{fig:llm-arch}, the basic attention mechanism in the Transformer is Multi-Head Attention (MHA), which employs separate, independent Key and Value vectors for each attention head.  
The number of heads for Key, Value, and Query in MHA is the same.
To mitigate the significant memory footprint caused by MHA, modern state-of-the-art LLMs, such as Llama-3~\cite{24arxiv-llama3} and Qwen2.5~\cite{24arxiv-qwen2.5}, have widely adopted Grouped-Query Attention~\cite{23emnlp-gqa}.  
GQA is a hybrid approach that strikes a balance between MHA and the extreme of Multi-Query Attention (MQA).  
As illustrated in Figure~\ref{fig:gqa}, in GQA, multiple query heads are grouped to share a single Key and Value head.  
This design drastically reduces the number of distinct K and V heads.

The architectural modification results in a significant reduction in the key-value cache. 
For instance, while an MHA-based model like OPT-66B~\cite{22arxiv-opt} uses 64 heads, a GQA-based model like Qwen2.5-70B~\cite{24arxiv-qwen2.5} uses only 8.  
This represents an eight times reduction in the volume of KV data per token.

\subsection{Re-Prefill with Prefix KV Cache}
\label{sec:reprefill}

% \textbf{Re-Prefill Phase.} 
The emergence of LLM applications, such as Retrieval-Augmented Generation~\cite{24arXiv-rag-survey, 24nips-uda, 24iclr-refusion, 25acl-seakr}, multi-turn conversations~\cite{25acl-f-mutli-turn, 25iclr-multi-turn, 25iclr-multi-turn-2, 25iclr-multi-turn-3}, long-context Question-Answering (QA)~\cite{25acl-longbench-v2, 24nips-babilong}, coding completion~\cite{21arxiv-codex, 25acl-code-complete, 24iclr-code-complete, 24icml-code-complete}, and multi-agent systems~\cite{hia, 24arix-stockagent, 24www-recai}, has fundamentally changed the serving workload. 
For RAG or long-context QA, one or multiple documents may serve as the context for dozens of user queries. 
In conversations, a lengthy system prompt or shared chat history serves as the common ground for multiple user sessions. 
Code completion often requires reading one or more source code files to understand the code logic and generate new code.
Multi-agent systems would reuse intermediate states, such as outputs of internal agents.
Among these applications, one key feature arises: they all create requests with long, shared prefixes.

Processing these shared prefixes from scratch for every request is computationally prohibitive and often results in significantly worse TTFT~\cite{24nips-sglang, 23sosp-vllm, 25asplos-flashgen, 25asplos-vattention}.
To alleviate this issue, existing offloading-based inference systems~\cite{24atc-as, 25fast-impress, 25eurosys-cacheblend, 25asplos-flashgen, 24osdi-infinigen} have proposed reusing the pre-computed KV cache, termed the shared prefix KV cache, to reduce redundant computations for subsequent requests.
We can serve new requests by loading the pre-stored KV cache for the shared prefix and then computing the KV cache for the new, unique suffix.
Unlike the typical prefill or decoding phase introduced in Section~\ref{sec:basics}, this process defines a distinct, critical phase, which we formally call the \textbf{Re-Prefill Phase}.

\begin{figure}[t]
\centering
  \includegraphics[width=1\linewidth]{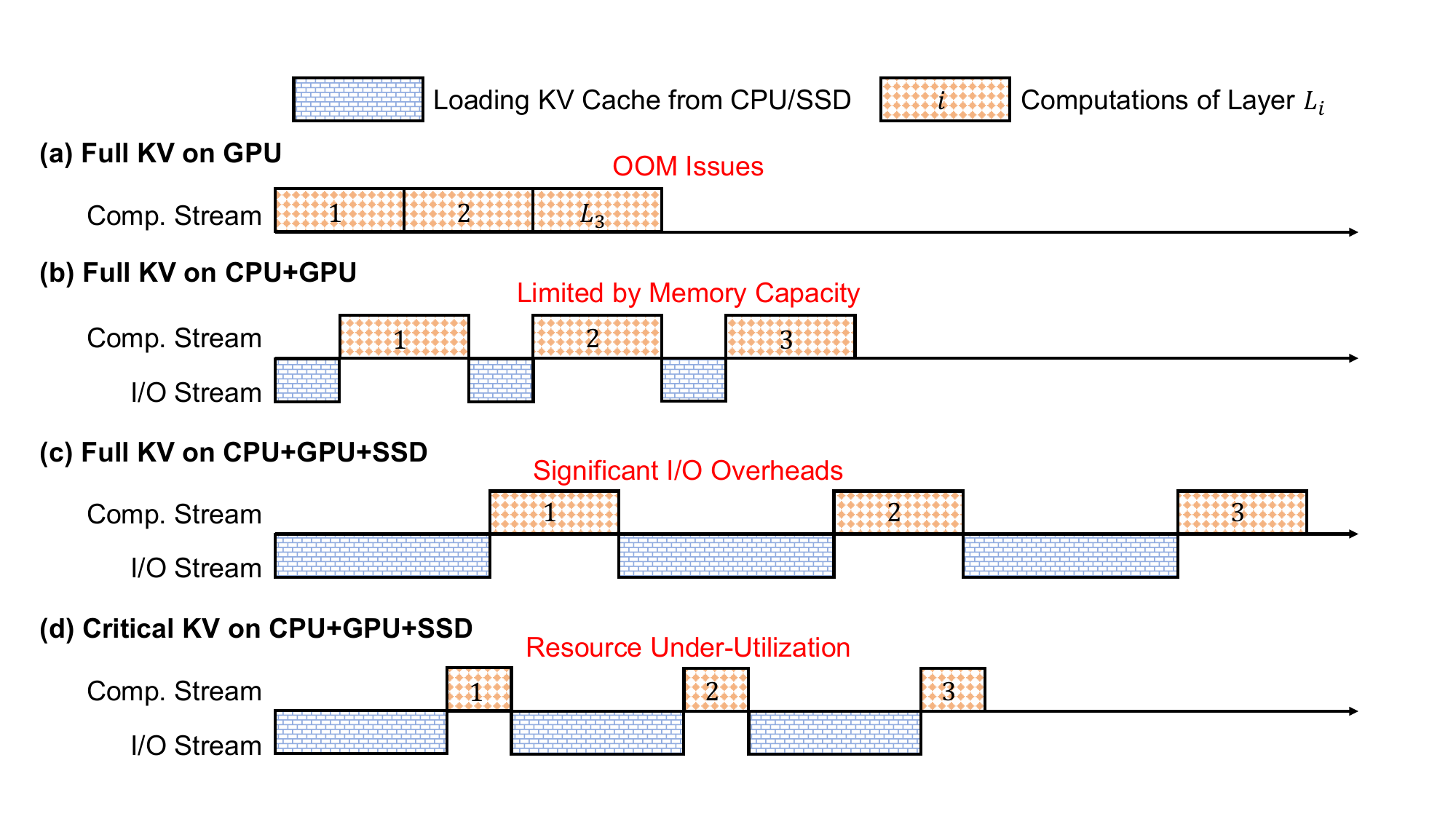}
  \caption{Paradigm of offloading-based inference systems.}
  \label{fig:existing-works}
\end{figure}

\begin{definition}
The \textbf{Re-Prefill Phase} is both computation- and I/O-intensive, where it processes a new request by: 
\begin{enumerate}[topsep=0pt,itemsep=-1ex,partopsep=1ex,parsep=1ex]
    \item Loading the pre-computed KV Cache of a shared prefix from a slower storage tier (e.g., CPU memory or SSD); 
    \item Computing the KV cache of the new, non-shared suffix tokens and performing the attention over the pre-computed KV cache to generate the first token. 
\end{enumerate}
\end{definition}

% Unlike the vanilla Prefill phase, which is predominantly compute-bound, the RePrefill phase presents a hybrid bottleneck. 
% Its performance is simultaneously constrained by the I/O bandwidth for loading the cached prefix and the computational overheads for processing the new tokens. 

The RePrefill phase has the following key properties that differentiate it from conventional LLM inference stages:
(1) \textbf{Intra-phase Sequential Dependency.} Within a single inference pass, only after loading the corresponding prefix KV cache from slower storage, we can then perform each layer's computation;
(2) \textbf{Resource Under-utilization.} The I/O operations and computations are not executed in parallel, although they are not conflicting. This leads to either I/O or compute resources remaining idle while the other is busy, severely hampering system efficiency.
% and (3) \textbf{Dynamic Bottleneck Shift.} If the amount of prefix KV cache size is larger, it would cause I/O-bound scenarios; while if the new input is longer, it would result in compute-bound constraints. The transitions between these two bottlenecks would limit the system's performance.
These intrinsic characteristics pose fundamental challenges that are not adequately addressed by existing serving systems.

\subsection{Offloading-based Inference Systems} 

\begin{figure}[t]
\centering
\includegraphics[width=1\linewidth]{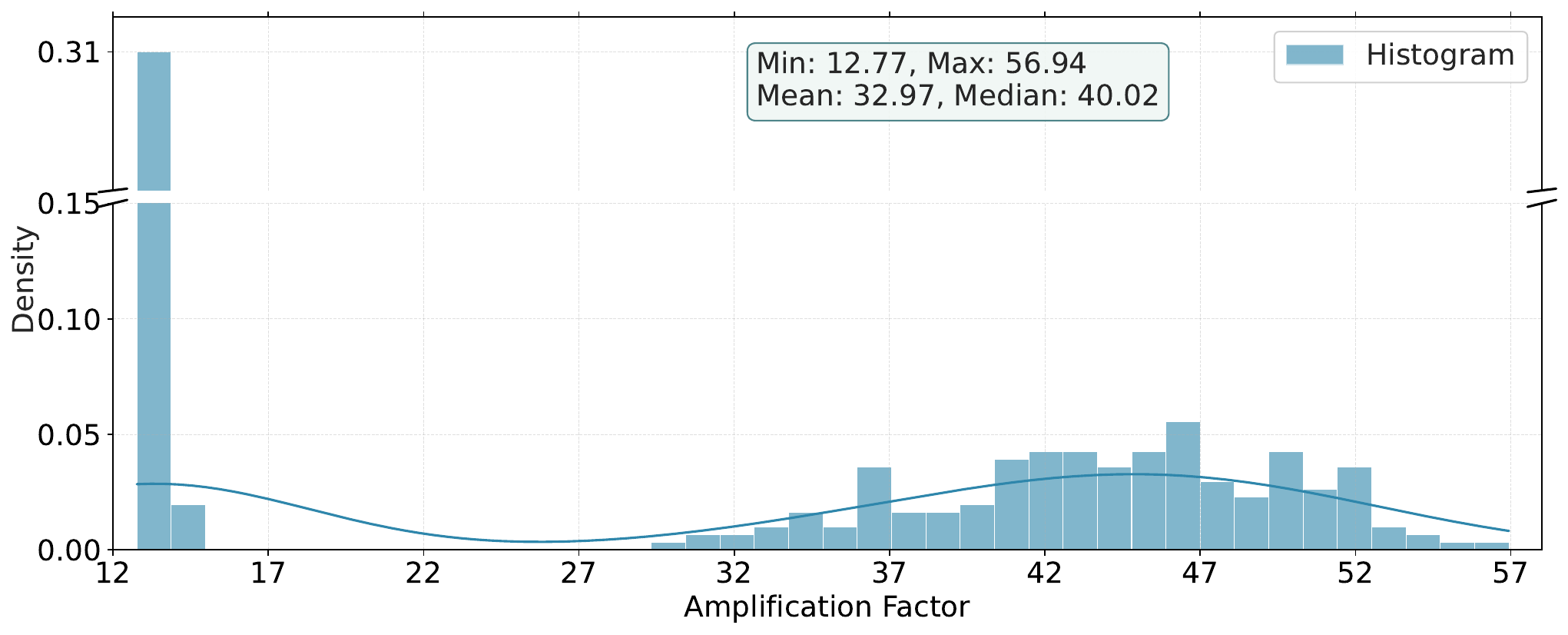}
% \begin{tabular}{c}
%   \subfigure[]{\includegraphics[width=1\linewidth]{figures/fig6-10.pdf}} \\
%   \subfigure[]{\includegraphics[width=1\linewidth]{figures/fig6-25.pdf}}
% \end{tabular}
  \caption{Read amplification of existing SOTA inference system (e.g., IMPRESS~\cite{25fast-impress}) when serving Qwen2.5-7B on the RTE dataset. The system adopts KV cache pruning methods to identify important KV caches, and the KV cache budget ratio is 25\%.}
  \label{fig:read-amplification}
\end{figure}

To accommodate the massive KV cache of long prefixes, modern LLM inference systems employ a multi-tier storage hierarchy comprising GPU memory, CPU main memory, and SSDs. 
We categorize them into four paradigms (Figure~\ref{fig:existing-works}),
\begin{itemize}[topsep=0pt,itemsep=-1ex,partopsep=1ex,parsep=1ex]
    \item \textbf{Full KV on GPU.} The default inference systems~\cite{24nips-sglang, 23sosp-vllm} used in deep learning inference usually store the entire KV cache exclusively in GPU memory. This system is often limited by the GPU memory capacity, which is infeasible for workloads with long shared prefixes.
    \item \textbf{Full KV on GPU + CPU.} To address out-of-memory issues, some inference systems~\cite{25sigmod-pqcache, 25acl-f-a2ats} extend the KV cache to CPU main memory, which serves as a secondary, larger cache tier. While this alleviates the memory capacity pressure, it still cannot adapt to the case where the KV cache grows during decoding.
    \item \textbf{Full KV on GPU + CPU + SSD.} For the long contexts that exceed total DRAM capacity, some systems~\cite{24atc-as, 25asplos-flashgen} offload most of the KV cache to SSDs. Although this provides near-unlimited storage, it introduces significant I/O overhead.
    \item \textbf{Critical KV on GPU + CPU and Full KV on SSD.} The most recent state-of-the-art (SOTA) systems~\cite{25fast-impress, 24osdi-infinigen} attempt to mitigate the I/O bottleneck by selectively loading only the important or critical KV pairs (or chunks), while keeping the full cache on SSD. This reduces the volume of data that needs to be transferred. However, they lack well-designed resource management methods, resulting in resource under-utilization.
\end{itemize}

This evolution highlights a persistent tension: expanding capacity introduces slower storage tiers whose I/O overheads will dominate the inference latency.
These limitations are particularly acute in the hybrid I/O-compute Re-Prefill phase, motivating the need for a storage system co-designed with the access patterns of modern, selective KV caching.
\section{Motivations and Challenges}
\label{sec:motivation}
\subsection{Motivations}

\begin{figure}[t]
\centering
    \includegraphics[width=1\linewidth]{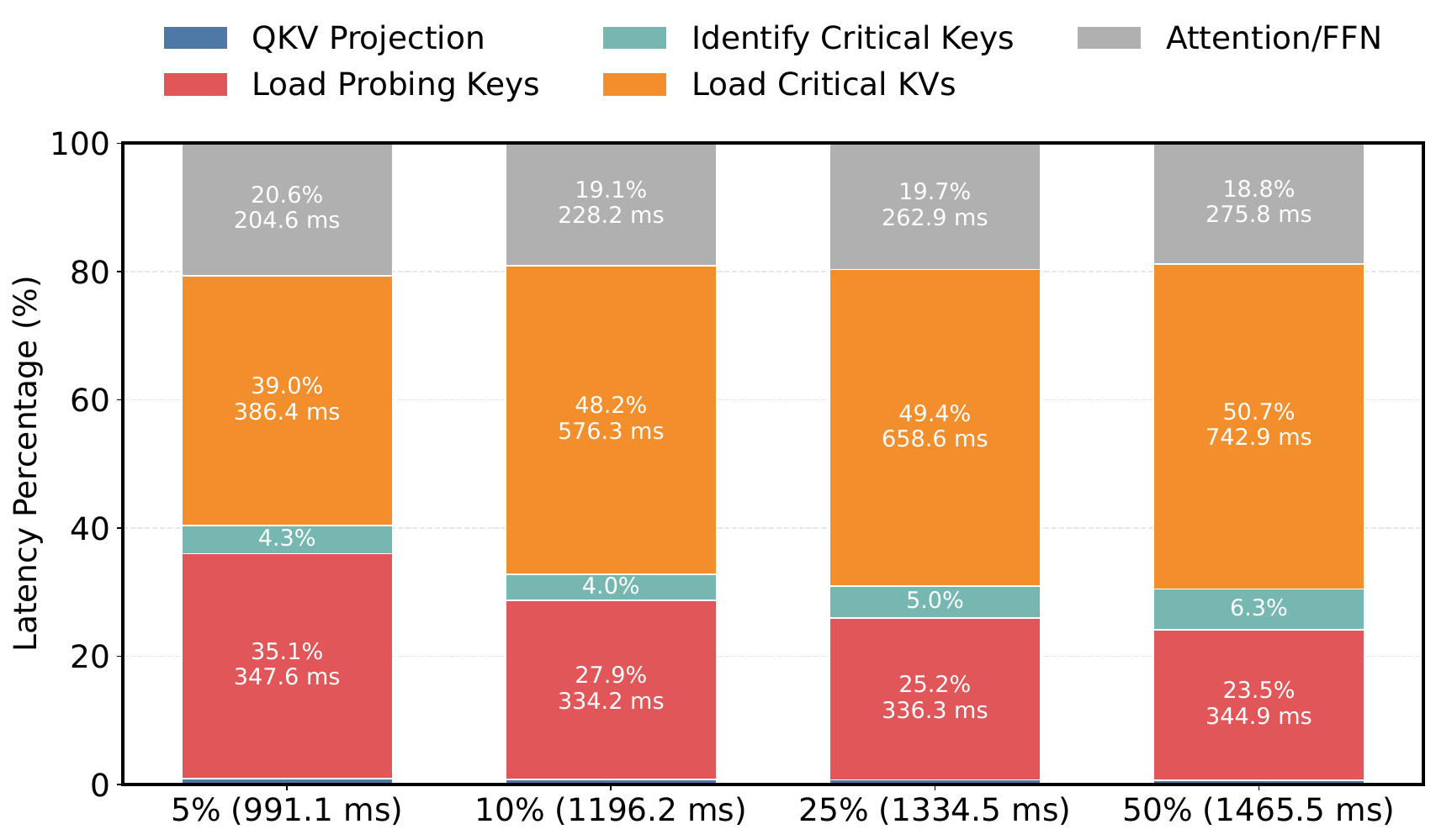}
    \caption{Latency breakdown of existing SOTA inference systems (IMPRESS~\cite{25fast-impress}) among different KV cache budget ratios.}
    \label{fig:breakdown}
\end{figure}

The Re-Prefill phase has emerged as a fundamental step in shared-prefix scenarios. 
% Its characteristics create unique challenges that are different from the vanilla prefill or decoding phases. 
% The efficiency of this phase is crucial for achieving high performance with a low TTFT in modern applications~\cite{25fast-impress, 24atc-as, 25eurosys-cacheblend}.
However, existing systems designed for LLM inference with shared prefix kv cache, such as AttentionStore~\cite{24atc-as} and IMPRESS~\cite{25fast-impress}, fail to address the unique characteristics of the Re-Prefill phase, especially when serving modern LLMs. 
Their limitations stem from outdated architectural assumptions and inefficient data management strategies.

\noindent\textbf{Motivation 1: Severe Read Amplification. }
The state-of-the-art offloading-based inference systems~\cite{25fast-impress, 24atc-as} usually operate with the KV cache at the granularity of chunks, which logically group multiple tokens into a chunk to align with I/O interfaces. 
To alleviate the memory overheads, IMPRESS adopts a sparse KV cache algorithm, such as H2O~\cite{23nips-h2o}, which reduces the number of KV caches required for attention operations.
This indicates that IMPRESS can only load those chunks that contain the required KV caches.
However, such a chunk-based approach will introduce significant read amplification when coordinating with a sparse KV cache algorithm.

% To understand the limitation, let's consider a typical configuration: 
% For the Qwen2.5-7B model, the KV cache for a single token (key or value only) occupies 28 KB (2B × 3584 × 4). 
% This already exceeds the typical SSD page size of 4KB~\cite{}. 
% When tokens are grouped into chunks (e.g., 16 tokens per chunk), each chunk holds several hundred KB of data. 
% If the importance identification mechanism selects only a few tokens within a chunk or a low budget ratio is adopted (e.g., 5\%), loading the entire chunk wastes I/O bandwidth and memory resources.
% Existing systems still operate at chunk granularity and thus suffer from this fundamental amplification.

We conducted a preliminary experiment to show the read amplification of IMPRESS~\cite{25fast-impress} when serving Qwen2.5-7B on the RTE dataset.
As shown in Figure~\ref{fig:read-amplification}, IMPRESS incurs about 12 times read amplification when serving most requests (about 31\%).
These read amplifications are introduced because only a small set of important tokens within the chunks is required.
Notably, the read amplification value could be extremely large  (from 25 to 56).
This happens when most of the KV cache prefix data for these requests is already cached, while the rest of the KV caches are distributed across multiple large-sized data chunks.
For example, a request requires loading an additional 11 tokens from the disk.
However, these tokens are stored across 9 chunks, resulting in a read amplification of 52.

\noindent\textbf{Motivation 2: Inefficient Resource Management and Under-utilization.}
As described in Section~\ref{sec:reprefill}, the Re-Prefill phase exhibits a strong intra-phase sequential dependency: each layer's KV cache of the shared prefix must be loaded before each layer's computation can proceed.
For example, identifying critical tokens requires loading prefix key vectors, or performing attention operations requires loading the prefix KV caches. 
Although existing systems~\cite{25fast-impress, 24atc-as} can overlap these I/O and compute steps to a limited extent, either the GPU or the I/O subsystem is still idle for significant periods. 
% The root reason for the limited overlapping is that these systems address the problem only from a system-level perspective, seeking to pipeline fixed stages of I/O and computation. 
% They overlook the algorithmic characteristics inherent to the model itself, which can reduce some unnecessary computations or I/Os.

We conducted a preliminary experiment to break down the latency of IMPRESS in Figure~\ref{fig:breakdown} and quantify the severe resource under-utilization. 
Across various KV cache budget ratios (5\%-50\%), the I/O-intensive stages, including loading the probing keys and loading the critical KVs, constitute over 65\% of the Re-Prefill phase, while the compute-intensive stages account for less than 35\%.
This imbalance confirms that the GPU remains idle for the majority of the time, waiting for KV data to be loaded. 
Even with some overlapping enabled, the fundamental sequential dependency forces I/O to be the critical path.

\begin{figure}[t]
\centering
  \includegraphics[width=0.8\linewidth]{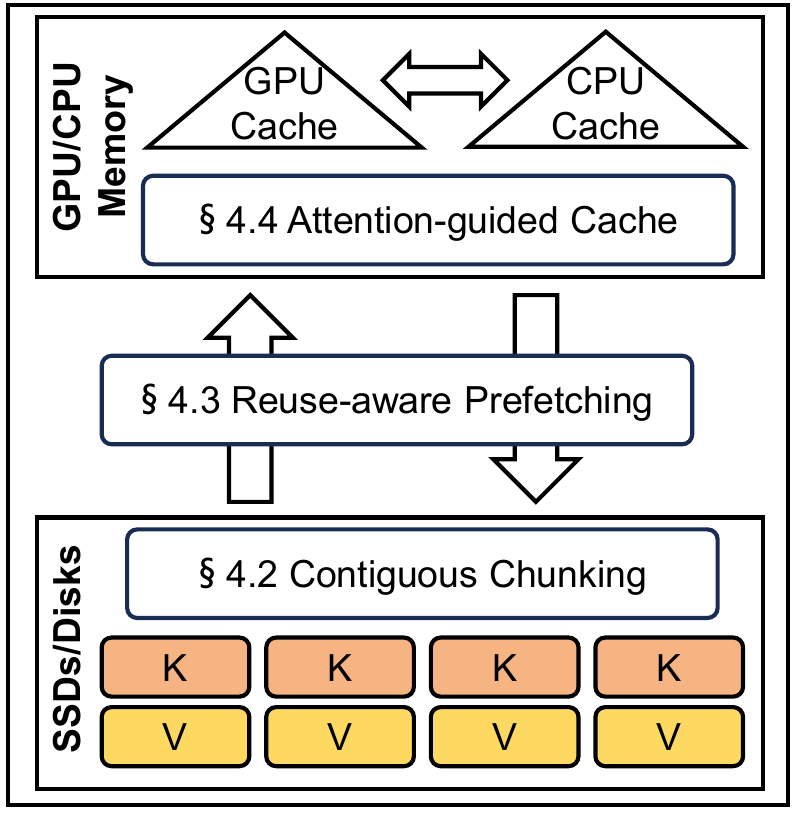}
  \caption{Overview of ContiguousKV Design.}
  \label{fig:overview}
\end{figure}

\subsection{Challenges}
\label{sec:challenge}
The above motivations translate into two concrete design challenges for building an efficient offloading-based inference system for the Re-Prefill phase.

\noindent \textbf{Challenge 1: Exploiting Hardware Capability to Eliminate Read Amplification without Sacrificing Bandwidth. } 
Modern storage devices (e.g., NVMe SSDs) support fine-grained random reads, which can, in principle, eliminate the read amplification caused by loading large chunks containing sparse important data. 
However, naively issuing a massive number of fine-grained, non-sequential read requests would serialize I/O operations and fail to utilize the high sequential bandwidth of the underlying device. 
The key challenge lies in reorganizing data and designing an access mechanism that simultaneously enables fine-grained loading of important data while preserving large, sequential read streams to maintain high I/O throughput.

\noindent\textbf{Challenge 2: Enabling Effective Prefetching Beyond the Sequential Dependency.} 
The root cause for the severe resource under-utilization is the intra-phase sequential dependency between compute and I/O of the Re-Prefill workload.
System-level optimizations that pipeline fixed I/O and compute stages for the current inference workflow have reached their limit.
Therefore, the key challenge is to leverage the LLM's internal characteristics to break this dependency, enabling effective prefetching ahead of time and maximizing I/O-compute overlap.

\section{ContiguousKV}
\label{sec:design}

\subsection{Overview}
\label{sec:overview}
We propose ContiguousKV to address the Re-Prefill bottleneck through a coordinated system design. Figure~\ref{fig:overview} presents the overview of ContiguousKV, which centers on three key innovations: the ContiguousChunk abstraction, a two-level prefetching engine, and an attention-guided cache controller. The ContiguousChunk abstraction resolves the granularity mismatch by creating a cohesive unit that aligns the data management granularity to the algorithmic granularity, eliminating wasteful data transfers where only specific tokens are needed (\S \ref{sec:fine-grained}). Our two-level prefetching mechanism exploits similarities to transform the sequential Re-Prefill process into a pipelined execution, effectively hiding I/O latency behind computation (\S \ref{sec:prefetch}). The attention-guided cache controller further optimizes memory utilization by prioritizing semantically important chunks based on the attention scores(\S \ref{sec:cache}).

\begin{figure}[t]
\centering
\begin{tabular}{ccc}
  \subfigure[Layers' similarity.]{\includegraphics[width=0.5\linewidth]{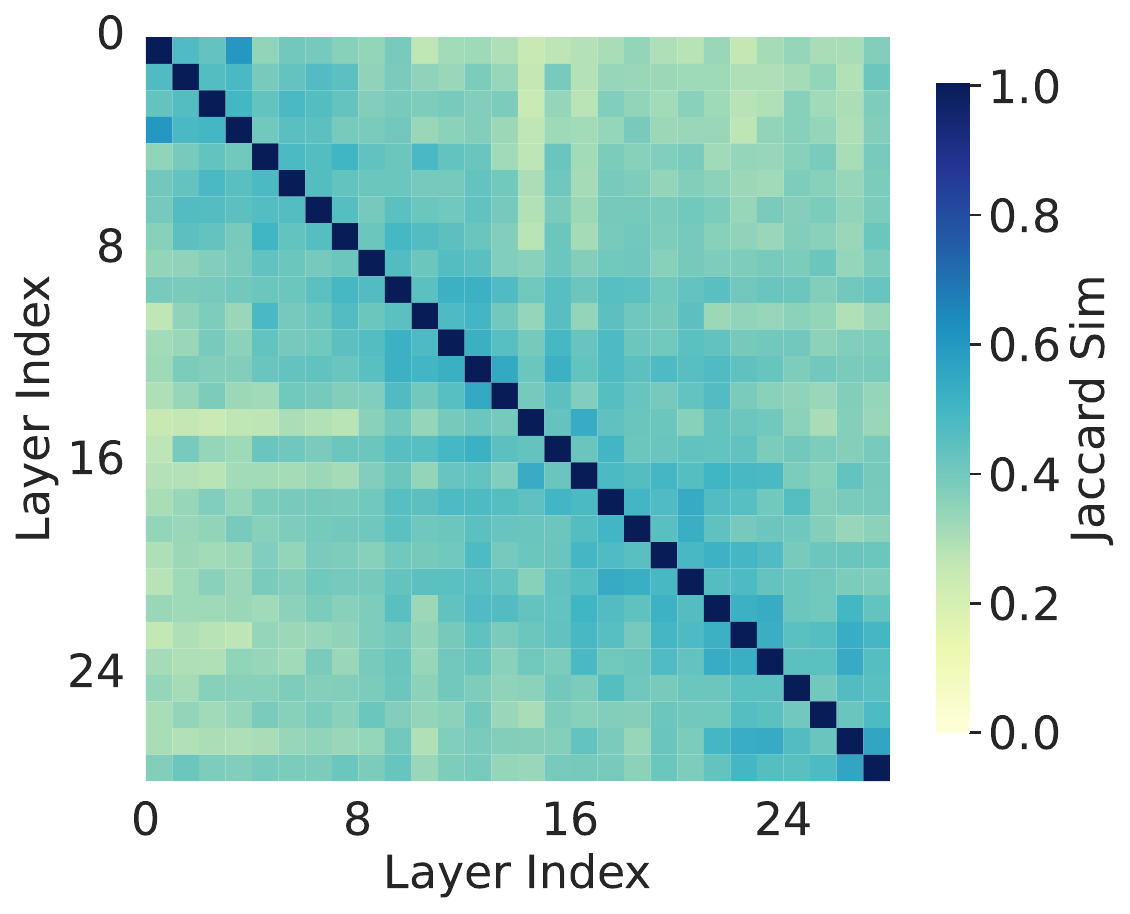}}
  \subfigure[Periods' similarity.]{\includegraphics[width=0.5\linewidth]{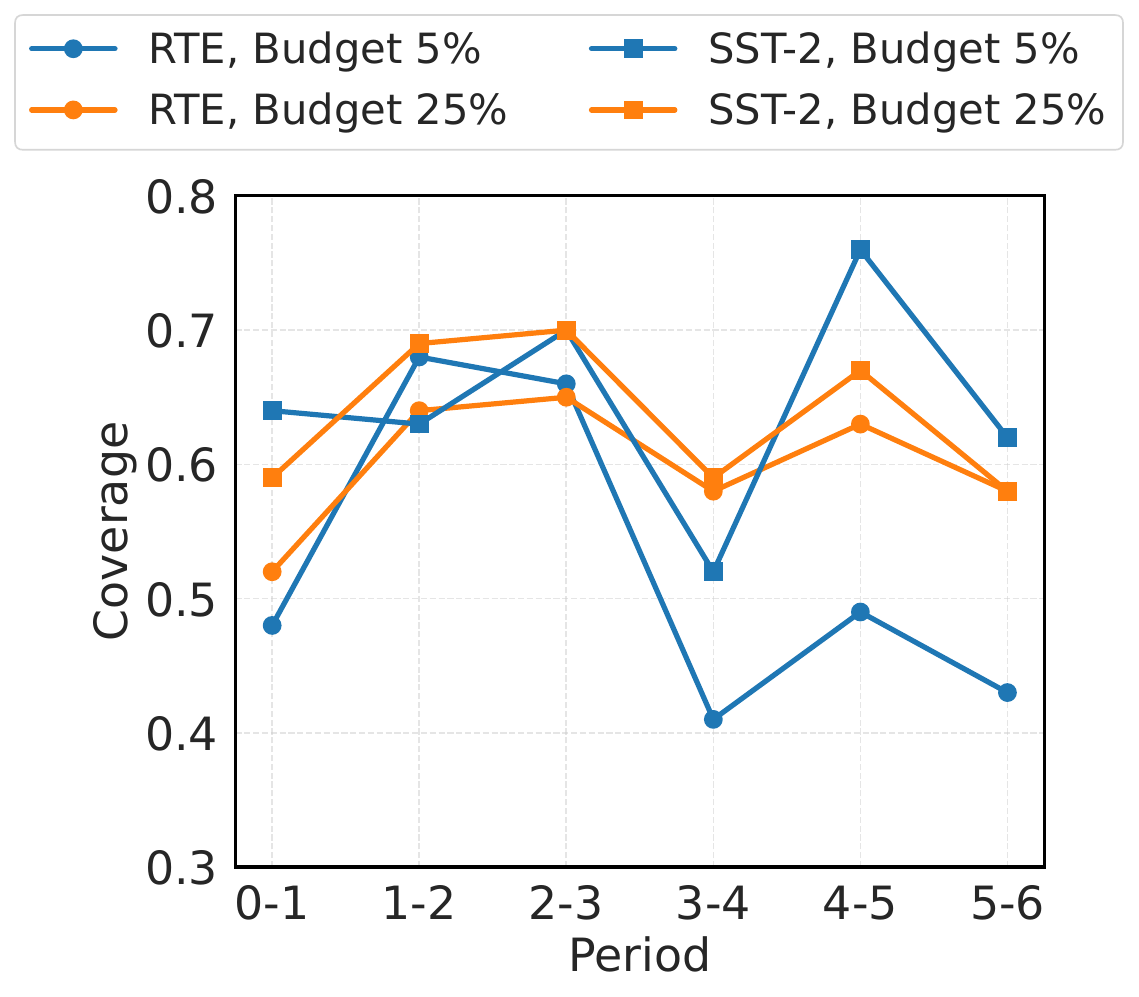}}
\end{tabular}
  \caption{The similarity between different layers' and different periods' important ContiguousChunk indices.}
  \label{fig:sim-period}
\end{figure}

\begin{figure*}[t]
\centering
  \includegraphics[width=0.8\linewidth]{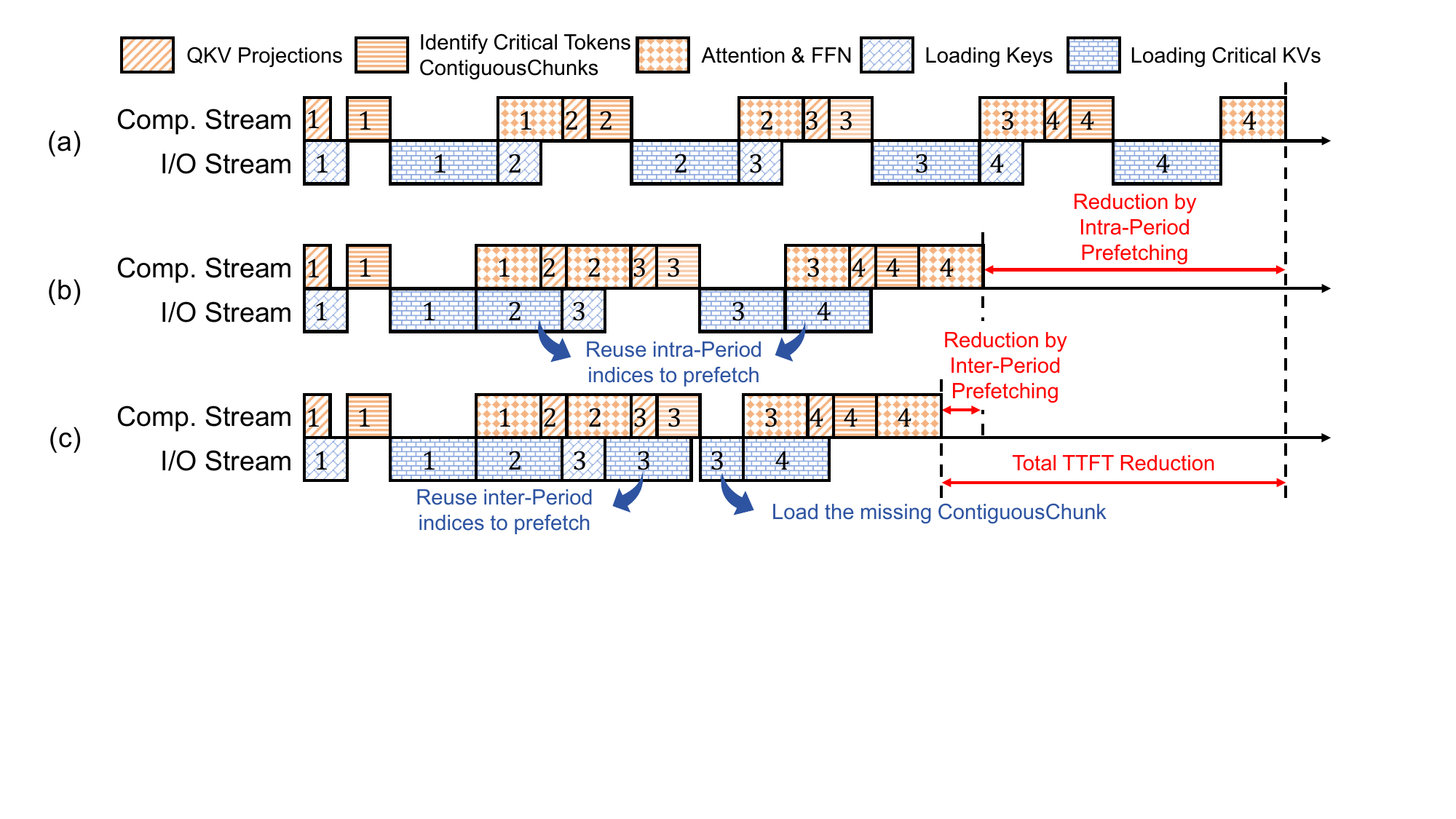}
  \caption{Comparisons of default prefetching in existing offloading-based inference systems and our two-level prefetching mechanism.}
  \label{fig:prefetch}
\end{figure*}

\subsection{Contiguously Chunking: Aligning System Granularity to Algorithm Granularity}
\label{sec:fine-grained}

KV cache management in offloading-based LLM inference faces a fundamental granularity mismatch between algorithmic decisions and system operations.
On the algorithmic side, token-level pruning methods~\cite{24nips-snapkv,23nips-h2o,23nips-scissorhands} select individual tokens based on attention scores but disrupt local semantic coherence in natural language.
Recent studies~\cite{25nips-chunkkv, 24acl-chunkattention} have demonstrated that combining consecutive tokens into groups can better preserve semantic context, leading to more effective KV cache reduction with higher output quality. 
On the system side, chunk-level offloading frameworks like IMPRESS~\cite{25fast-impress} manage KV cache in coarse data chunks (e.g., multi-megabyte blocks) to amortize I/O costs, but this introduces severe read amplification when only small portions of each chunk are needed (Figure~\ref{fig:read-amplification}).
We introduce \textbf{ContiguousChunk}, a unified granularity that aligns system I/O with algorithmic pruning to address performance bottlenecks.

% Most existing KV cache management techniques operate at the granularity of individual tokens~\cite{24nips-snapkv, 23nips-h2o, 23nips-scissorhands, 24iclr-streamingllm}. 
% While fine-grained, such token-level methods risk disrupting the local semantic coherence in natural language.  
% Recent studies~\cite{25nips-chunkkv, 24acl-chunkattention} have demonstrated that combining consecutive tokens into groups can better preserve semantic context, leading to more effective KV cache reduction with higher output quality. 
% To clearly distinguish this fundamental semantic unit from the data chunk used for I/O and caching in existing offloading-based inference systems, we introduce \textbf{ContiguousChunk}. 
\begin{definition}
    The \textbf{ContiguousChunk} is a set of consecutive tokens. %, which is the minimal granularity for performing KV cache management for preserving semantics.
    Formally, for an input prefix of length $n$, we partition it into $m = \lceil n / c \rceil$ contiguous ContiguousChunks, each containing up to $c$ tokens (e.g., $c=16$).
\end{definition}
% ContiguousChunk serves as the \textit{system unit} for data storage layout and I/O, to align with the algorithmic unit for importance scoring and eviction decisions, which preserves local semantics better than token-level granularity.
%; and (2)  eliminating read amplification inherent in coarse data chunks.

% The advantage of the ContiguousChunk granularity is its ability to preserve coherent semantic segments. 

% \textbf{System Granularity.}
% \subsubsection{Fine-grained Access with ContiguousChunk}
% \label{sec:layout}
% To address the severe read amplification inherent in prior chunk-based inference systems (Section~\ref{sec:challenge}), we introduce a fine-grained access mechanism operating at the granularity of ContiguousChunks. 
% While token-level random access could theoretically eliminate read amplification, it would generate an excessive number of small, non-sequential I/O requests.
% This will serialize operations and fail to utilize the high sequential bandwidth of modern SSDs. 
ContiguousChunk establishes a unified granularity that governs all critical operations: storage organization, cache management, and I/O prefetching. This consistent granularity eliminates the inefficiencies inherent in systems that employ different granularities across different layers.
Unlike systems like IMPRESS that manage storage in coarse data chunks but must handle eviction and selection at finer (e.g., token) granularity, our design ensures that the unit of eviction, the unit of storage, and the unit of prefetching are all at the ContiguousChunk granularity. 
This alignment is fundamental to our efficiency.

% Unlike IMPRESS's coarse data chunks that force loading of irrelevant KV pairs, our system stores each ContiguousChunk contiguously on SSD and fetches only the selected ContiguousChunks.
% This eliminates read amplification.

Consider a typical configuration:
For the Qwen2.5-7B model, the KV cache for a single token occupies approximately 28 KB (3584 hidden dimensions × 4 heads × 2 Bytes per float).
This already exceeds typical SSD page sizes (4KB).
When tokens are grouped into coarse chunks, e.g., 64 tokens per chunk as in prior systems~\cite{25fast-impress, 24atc-as}, each chunk holds about 1,800 KB of KV data.
If the importance selection mechanism identifies only a few key tokens (e.g., 16 tokens) within the chunk (or if a low budget ratio like 5\% is adopted), loading the entire data chunk may lead to a read amplification ratio of 4.
In contrast, the ContiguousChunk granularity aligns system I/O with algorithmic pruning by setting the chunk size to 16 tokens. 
Consequently, fetching one important ContiguousChunk following chunk-based KV cache pruning algorithms would result in zero read amplification. 
This alignment ensures that nearly every byte read from storage is utilized by the computation, thereby eliminating bandwidth waste and maximizing I/O efficiency.

\subsection{Reuse-aware Asynchronous Prefetching}
\label{sec:prefetch}
In this section, we will introduce reuse-aware asynchronous prefetching techniques.
We first show the similarity across different layers when adopting a fixed, periodic reuse strategy.
Then, we introduce the prefetching strategies within or between periods.

\subsubsection{Similarity in Critical ContiguousChunk Indices}
\label{sec:reuse}

Similar to previous work~\cite {25nips-chunkkv, 24acl-chunkattention}, we found that the set of important ContiguousChunk indices remains highly similar across Transformer layers.
Thus, we adopt the same strategy to simply group a set of consecutive layers, termed as \textbf{Periods}, for sharing the same ContiguousChunk indices.
\begin{definition}
    The \textbf{Period} refers to a set of consecutive layers, where they share the same important ContiguousChunk indices. Each Period contains $p$ layers (e.g., $p=4$).
\end{definition}
The important ContiguousChunk indices used for each Period are determined during the first layer of the Period, where the identification process is described in Section~\ref{sec:fine-grained}.
For the rest layers within the same Period, we reuse the same ContiguousChunk indices.

\noindent\textbf{Similarities between Periods.}
Figure~\ref{fig:sim-period}(a) shows the similarities between layers, where we can observe that adjacent layers tend to have similar indices, while far layers are different.
% is no obvious constant periodic pattern between layers.
% Simple organizing layers into fixed Periods ignores the similarities between layers' indices in different Periods.
We conducted an observation experiment on Qwen2.5-7B in Figure~\ref{fig:sim-period}(b) to demonstrate that the important ContiguousChunk indices from different periods also exhibit high similarities.
We use the coverage ratio, which is calculated by the number of the same ContiguousChunk indices between adjacent Periods, to represent the similarity.
The results show that the similarity between Periods ranges from 52\% to 64\% on average.
This demonstrates great potential to also reuse parts of the ContiguousChunk indices between Periods.
Taking advantage of the high similarities, we will introduce prefetching mechanisms to hide the I/O costs during computation. 

% Figure~\ref{fig:sim-period} shows that the adjacent layers tend to have high similarity in the important token indices.
% Although there is no obvious constant periodic pattern between layers, we can still simply organize layers into fixed, consecutive \textbf{Periods}.
% Within each period, the important ContiguousChunk indices are identified once at the period's start and are reused for all subsequent layers. 

% This fixed-period reuse strategy has been demonstrated to be effective in improving model performance in previous works~\cite{25nips-chunkkv, 24acl-chunkattention}.
% But more critically, this fixed-period reuse strategy creates a regular and predictable access pattern for the storage subsystem. 
% Unlike adaptive schemes where reuse boundaries vary per request, our fixed periods allow the system to confidently aggregate, reorganize, and prefetch data corresponding to all layers within a known boundary. 
% This regularity fundamentally enables our Period-First data layout (Section~\ref{sec:layout}) and our two-level asynchronous prefetching mechanism (Section~\ref{sec:prefetch}), which together are key to mitigating the I/O bottleneck and resource under-utilization in the Re-Prefill phase.

\subsubsection{Intra/Inter-Period Prefetching Mechanism}
\label{sec:prefetch-steps}

Figure~\ref{fig:prefetch}(a) illustrates the execution timeline of the Re-Prefill phase in existing systems using a four-layer LLM as an example.  
The computation for each layer follows a strict, sequential dependency. 
First, the system loads the prefix's key vectors from SSD; concurrently, it performs the QKV projection to obtain the query vectors for the current layer. With both queries and prefix keys in GPU memory, it identifies the critical tokens for this layer. Finally, it loads the corresponding KV cache for these tokens to complete the attention computation.
Although some prefetching of the next layer's prefix keys is possible after critical token identification~\cite{25fast-impress,24atc-as}, the fundamental dependency, that the KV cache for layer $l$ cannot be loaded until the critical tokens for layer $l$ are identified, creates significant idle periods for both GPU and I/O resources. 
These idle ``bubbles'' serialize I/O and compute, underutilizing system resources and limiting throughput.

To fundamentally break this sequential dependency and reclaim idle bubbles, we design a combined intra- and inter-period prefetching strategy that leverages reusable critical ContiguousChunk indices across layers, as described in Section~\ref{sec:motivation}. 
The core idea is to orchestrate asynchronous I/O operations to run ahead of computations at two distinct granularities: within a Period (Intra-Period Prefetching) and across consecutive Periods (Inter-Period Prefetching). 
This pipeline effectively hides I/O latency, minimizes idle bubbles, and maximizes concurrent utilization of I/O and compute resources.

% As described in Challenge 1 (Section~\ref{sec:challenge}), the Re-Prefill phase exhibits a strong sequential dependency: the KV cache of the shared prefix must be loaded before each computation (e.g., probing critical tokens or attention operations) can proceed. 
% While some I/O operations can be overlapped with computations, this dependency forces existing systems to execute these steps in a strictly serial manner, leaving either the GPU or the I/O subsystem idle most of the time and creating idle bubbles in the timeline. 
% This issue is limited by the inference scheme, which cannot be addressed only by system optimizations.

\noindent\textbf{Intra-Period Prefetching: Overlapping I/O with Compute.}
By reusing the same ContiguousChunk set, our intra-period prefetching mechanism prefetches the critical ContiguousChunk to pipeline I/O operations with layer computations, as illustrated in Figure~\ref{fig:prefetch}(b).
Taking an example of a 4-layer LLM, the processes for a Period of 2 layers are:
\begin{enumerate}[topsep=0pt,itemsep=-1ex,partopsep=1ex,parsep=1ex]
    \item At the first layer ($L_1$) within a Period, the system performs the standard steps: loading prefix keys, computing queries, and identifying the critical ContiguousChunk indices. Crucially, this index set is determined once and reused for all subsequent layers in the same Period.
    \item Immediately after this identification, the system issues \textit{asynchronous I/O requests} to load the corresponding KV cache for all critical ContiguousChunks across all layers within the same Period. Notably, the loading process for different layers is sequential, which can be pipelined with each layer's computations.
    \item While the KV cache for the layer $L_1$ is being fetched from storage, the computation for $L_1$ proceeds.
    \item For each subsequent layer ($L_2$) in the Period, its required KV cache has likely already been prefetched into GPU memory, eliminating the per-layer I/O wait. The prefetching of the KV cache for later layers is pipelined with the computation of earlier layers.
\end{enumerate}

% Although intra-period prefetching can mitigate most idle bubbles within each Period, idle bubbles during the computation of the first Layer of each Period still cannot be avoided.
% This is because the first layer's critical ContiguousChunks' indices must be determined after loading the required keys or values, and the first layer's computations must start after loading the critical ContiguousChunks.
% But benefited from the observations in Section~\ref{sec:reuse}, we can extend the intra-Period complete reuse to the inter-Period partial reuse.

\noindent\textbf{Inter-Period Prefetching: Warming Up Period Boundaries.}
While intra-period prefetching eliminates bubbles between layers within the same Period, a significant bubble remains at the start of each Period where the first layer must load its prefix keys and compute before any identification can occur. 
To mitigate this, we leverage the observed similarity in critical ContiguousChunk indices between consecutive Periods (see Figure~\ref{fig:sim-period}).
We employ a lightweight, speculative prefetching strategy, illustrated in Figure~\ref{fig:prefetch}(c):
\begin{enumerate}[topsep=0pt,itemsep=-1ex,partopsep=1ex,parsep=1ex]
    \item Before a Period (e.g., Period $i$ but except for the first Period) officially begins its computation, we asynchronously prefetch the KV cache for the critical ContiguousChunks of the previous Period ($i-1$) for the upcoming layers in Period $i$.
    \item Once Period $i$ starts and its first layer definitively identifies its own critical ContiguousChunk set, the system only needs to load the difference, the missing ContiguousChunks not already prefetched. This results in a much shorter, final loading step.
\end{enumerate}
% To further mitigate idle bubbles, we propose inter-period prefetching. 
% According to Figure~\ref{fig:sim-index} in Section~\ref{sec:reuse}, although we group layers into Periods following the fixed, consecutive rules, the critical ContiguousChunks' indices also exhibit relatively high similarities between Periods.
% Therefore, as shown in Figure~\ref{fig:prefetch} (c), after loading the prefix keys of the next Period, we can reuse the last Period's critical ContiguousChunks' indices to load some potentially important ContiguousChunks of the current Period into memory.
% Once the actual important ContiguousChunks' indices are ready, we then load the missing ContiguousChunks. 
% It is a shorter loading process as most ContiguousChunks have been loaded in advance. 

% This inter-period prefetching can significantly mitigate the idle bubbles between Periods except for the first Period.
% It further reduces the TTFT slightly, maximizing both compute and I/O resource utilization.
This inter-period prefetching acts as a ``warm-up'', significantly reducing the initial I/O bubbles. 
Together, our intra- and inter-period prefetching mechanism ensures near-continuous utilization of both GPU compute and I/O bandwidth, transforming the Re-Prefill phase from a sequential bottleneck into a highly pipelined and efficient process.

\subsection{Attention-based Cache Management}
\label{sec:cache}

While the intra/inter-Period prefetching mechanism effectively reduces I/O latency and idle time, the finite capacity of GPU and CPU memory necessitates intelligent cache management to maximize the hit rate for critical KV data. 
Existing cache policies in offloading-based systems, such as LRU, LFU, or even the score-based policy in IMPRESS~\cite{25fast-impress}, rely on coarse-grained statistics like access frequency and static importance ratios. 
They fail to capture the nuanced, query-dependent semantic importance of cached data as reflected by the model's own attention mechanism.

To address this, we introduce an attention-aware cache policy that manages ContiguousChunks based on an attention-aware cache score.
It computes the cache score $S_j$ of the ContiguousChunk $j$ based on its contribution to model inference (quantified by its attention score $A_j$) and its access frequency $F_j$.
This cache score can help the cache manage ContiguousChunks in a holistic, semantically-informed manner.

Similar to existing chunk-based KV cache selection methods~\cite{25nips-chunkkv}, we compute the importance score of each ContiguousChunk by accumulating the importance scores of all tokens within it.
For example, given an input, we first compute the token-level attention score $h_{qk}=softmax(h_{q}\cdot h_{k})$ using the standard attention mechanism, where $h_{qk}\in \mathbf{R}^{n\times n}$.
Then, we obtain the important score $a_i$ of $i$-th token by summing $h_{qk}$ over the second dimension.
The importance score $A_j$ for the $j$-th ContiguousChunk is the sum of the attention scores of $c\cdot(j-1)$-th to $c\cdot j$-th tokens:
\begin{equation}
\label{eq:score}
    A_j=a_{c\cdot (j-1)}+a_{c\cdot (j-1)+1} +\cdots + a_{c\cdot j- 1}
\end{equation}
% After obtaining all importance scores, all ContiguousChunks are then ranked according to their scores $A^l_j$.
% Finally, important ContiguousChunks are selected with the top-$k$ highest scores $A^l_j$.

For each ContiguousChunk $j$ stored in the cache, we maintain two dynamic values:
\begin{itemize}[topsep=0pt,itemsep=-1ex,partopsep=1ex,parsep=1ex]
    \item \textbf{Cumulative Attention-based Importance $I_j$.} The initial value of $I_j$ is zero. After processing a request, we record the attention score $A_j$ for each critical ContiguousChunk $j$ involved in the computation, calculated in Equation~\ref{eq:score}. Then, we update $I_j$ by adding the newly calculated ContiguousChunk's attention score $A_j$, i.e., $I_j=I_j+A_j$.
    \item \textbf{Access Frequency ($F_j$).} We track the number of times ContiguousChunk $j$ has been accessed (loaded into the GPU memory or CPU memory). 
\end{itemize}
Then, the cache score is the product of its importance and frequency:
\begin{equation}
    S_j=I_j\times F_j
\end{equation}

Based on this cache score, we utilize two min-heaps to manage the ContiguousChunks on both the GPU and CPU.
Specifically, at the initialization stage of the inference system, we create two empty min-heaps in the CPU memory.
During inference, before each prefetching process starts, we use the identified important ContiguousChunk indices to check if they are already cached.
For those not in the cache, we dispatch I/O threads to load them from SSD, as illustrated in Section~\ref{sec:prefetch-steps}. 
Once loaded, these ContiguousChunks are transferred to the GPU for computation, and their cache scores are updated simultaneously. 
According to their scores, they are then inserted into either the GPU heap or the CPU heap. 
Both heaps will evict low‑scored ContiguousChunks. 
Notably, ContiguousChunks evicted from the GPU heap are either inserted into the CPU heap (if their scores remain relatively high) or evicted from memory entirely. 
We store all cache scores, including those ContiguousChunks that are evicted from memory, in an in-memory table.

\subsection{Implementation Details}
% \noindent\textbf{Code Design and Implementations.}
As none of the existing state-of-the-art offloading inference systems (e.g., IMPRESS, AttentionStore) are open-sourced, we build ContiguousKV upon a well-known, publicly available LLM offloading inference framework, FlexGen~\cite{23icml-flexgen}. 
We extensively modify its core implementation to accommodate modern LLM architectures (e.g., Qwen2.5 series) and to integrate novel KV cache selection algorithms (e.g., H2O~\cite{23nips-h2o}, ChunkKV~\cite{25nips-chunkkv}). 
Specifically, we implemented two key classes: \textit{InferenceEngine} for system granularity alignment and two-level prefetching mechanisms, and \textit{CacheManager} for our attention-guided cache management.

% \noindent\textbf{System Configuration and Parameters.}
Our implementation carefully manages system resources to maximize efficiency without introducing significant overhead.
First, the data prefetched asynchronously does not occupy additional memory space beyond the preset memory.
Instead, we partition the given memory budget (e.g., 10 GB) into dedicated cache areas and small prefetch buffers.
In our experiments, within a 10 GB total memory constraint, we allocate only 0.2 GB and 0.4 GB as prefetch buffers on the GPU and CPU, respectively.
Second, to control the overlap between I/O and computation and maximize resource utilization, we introduce a key parameter, the subperiod\_size, which determines how many layers' KV cache must be loaded before the computation for the current Period can begin.

\begin{table}[t]
\caption{Detailed configurations of prefixes.}
\label{tab:conf}
\begin{center}
\begin{tabular}{lccccc}
\toprule
\multirow{2}{*}{\textbf{Datasets}} & \multirow{2}{*}{\textbf{\# Examples}} & \multirow{2}{*}{\textbf{Length}} & \multicolumn{3}{c}{\textbf{Size (GB)}} \\
& & & 7B & 14B & 32B \\
\midrule
\textbf{SST-2} & 100 & 3.8k & 67 & 226 & 301 \\
\textbf{SubJ} & 110 & 4.4k & 67 & 230 & 305 \\
\textbf{TREC} & 120 & 5k & 75 & 257 & 343 \\
\textbf{RTE} & 80 & 6k & 68 & 233 & 311 \\
\bottomrule
\end{tabular}%
\end{center}
\end{table}

\begin{figure*}[t]
\centering
\begin{tabular}{cc}
  \subfigure[SST-2]{\includegraphics[width=0.5\linewidth]{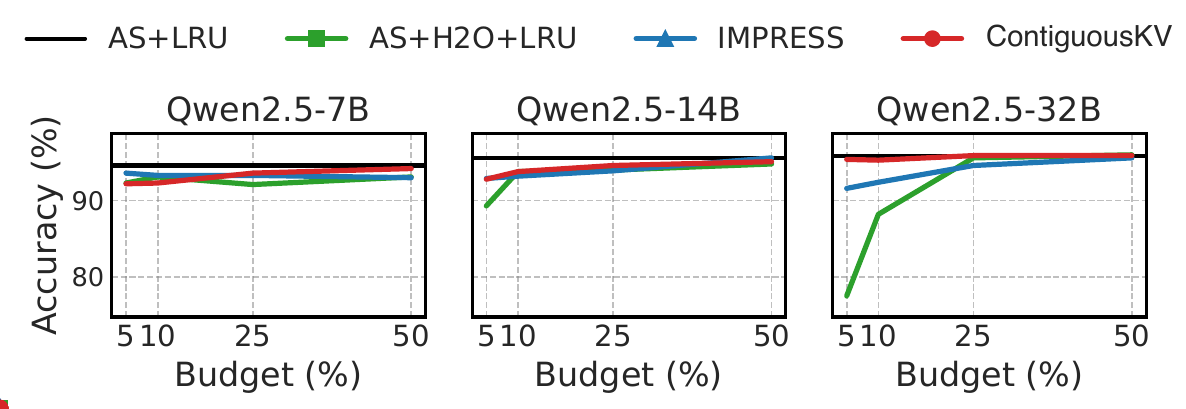}}
  \subfigure[SUBJ]{\includegraphics[width=0.5\linewidth]{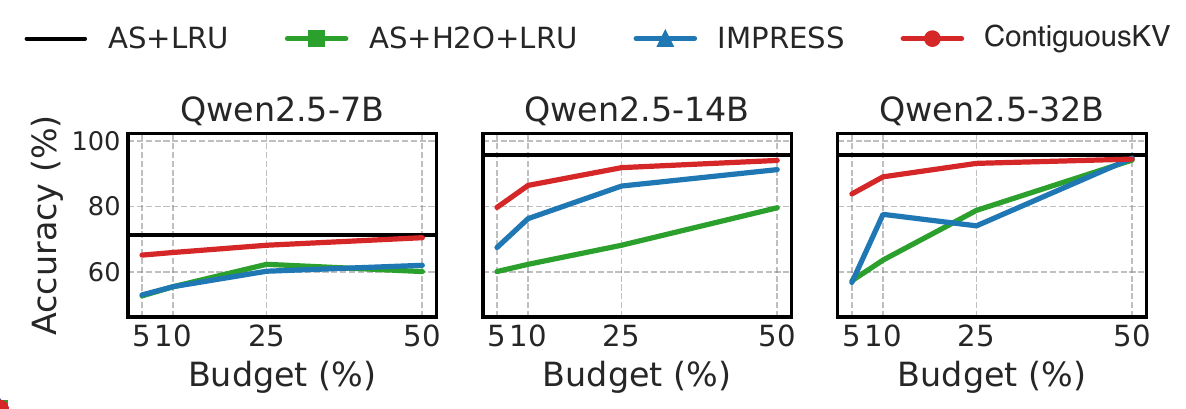}} \\
  \subfigure[TREC]{\includegraphics[width=0.5\linewidth]{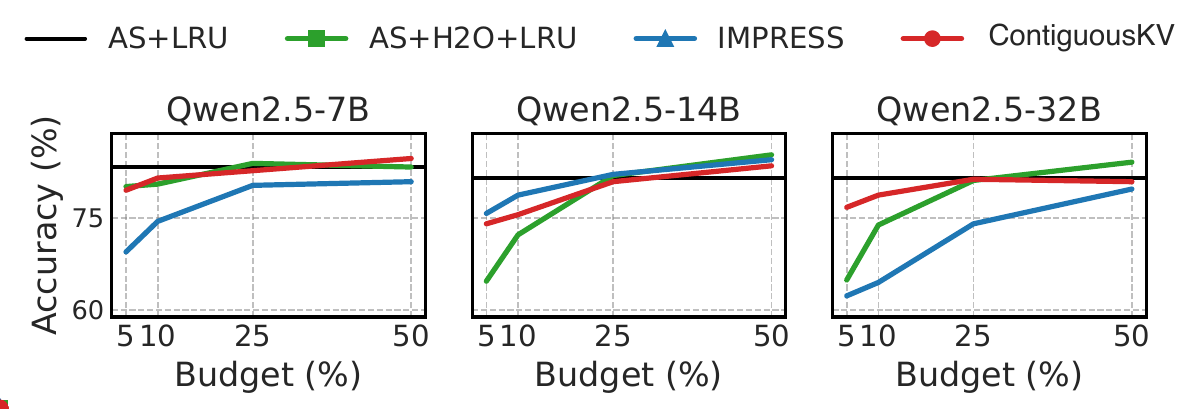}}
  \subfigure[RTE]{\includegraphics[width=0.5\linewidth]{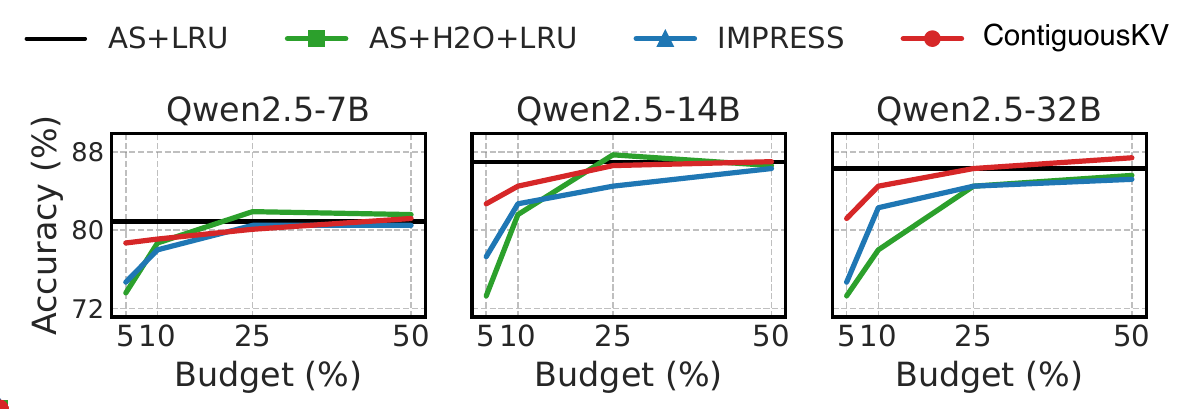}}
\end{tabular}
  \caption{Accuracy of different inference systems across four datasets and three LLMs.}
  \label{fig:acc}
\end{figure*}

\section{Evaluation}
\label{sec:evaluation}
\subsection{Experimental Setup. }

\noindent\textbf{Models and System Configuration.}
We evaluate ContiguousKV using the Qwen2.5 series of models~\cite{24arxiv-qwen2.5} over three scales: Qwen2.5-7B, Qwen2.5-14B, and Qwen2.5-32B. 
All experiments are conducted on a server equipped with 2x Intel Xeon Platinum 8370C CPUs (64 cores), 128 GB of DRAM, one NVIDIA A800 GPU with 80 GB HBM2e memory, and a 4 TB Samsung 990 Pro NVMe SSD. The measured peak read bandwidth of the SSD is about 7.45 GB/s.The GPU is connected to the host via PCIe 4.0 ×16 with a measured bidirectional bandwidth of about 32 GB/s.

\noindent\textbf{Datasets and Metrics.}
We select four representative natural language understanding tasks to simulate the Re-Prefill phase: SST-2~\cite{13emnlp-sst-2} for sentiment analysis, Subj~\cite{04acl-subj} for subjectivity classification, TREC~\cite{00sigir-trec} for question classification, and RTE~\cite{09tac-rte} for textual entailment. 
We evaluate LLMs on these tasks in a prompt-based, few-shot manner~\cite {24iclr-refusion, 20acl-lmbff}.
These datasets are well-suited for evaluating the Re-Prefill phase as they are formulated as prompt-based generation tasks that typically require the model to generate only a single token or a very short sequence (e.g., a label word). 
This characteristic minimizes the impact of the decoding phase on end-to-end latency, ensuring that our performance measurements (TTFT) primarily reflect the efficiency of the shared prefix processing and the associated I/O-compute orchestration, i.e., the core focus of this work. 
We measure results in accuracy (as in prior works~\cite{20acl-lmbff, 24iclr-refusion}), average TTFT, and Tail Latency (P95 TTFT).
We conduct experiments under four different KV cache budget ratios, i.e., 5\%, 10\%, 25\%, and 50\%.

\noindent\textbf{Baseline Inference Systems.}
We compare ContiguousKV against state-of-the-art offloading-based inference systems designed for the Re-Prefill phase:
\begin{itemize}[topsep=0pt,itemsep=-1ex,partopsep=1ex,parsep=1ex]
    \item \textit{AS+LRU}~\cite{24atc-as}: AttentionStore (AS) serves LLM inference with full prefix KV caches over a multi-tier storage system.
    \item \textit{AS+H2O+LRU}: We follow the baseline used in prior work~\cite{25fast-impress}, which combines AS+LRU with important KV selection algorithms, e.g., H2O~\cite{23nips-h2o}. Unlike AS+LRU, this baseline only loads important tokens into memory rather than full prefix KV caches.
    \item \textit{IMPRESS}~\cite{25fast-impress}: The current state-of-the-art offloading-based inference system, which selectively loads partial keys for importance identification and employs a score-based cache policy.
\end{itemize}
Since none of these baselines are open-source, we reimplement them to the best of our ability based on their papers.
For AS+LRU and AS+H2O+LFU, we disable their scheduler-aware optimizations to make them more suitable for general scenarios, such as preemptive scheduling environments, which aligns with IMPRESS's paper~\cite{25fast-impress}.

\noindent\textbf{Configurations.}
We construct shared prefixes by sampling data from each category of the training dataset as few-shot examples.
The sampling process satisfies a normal distribution.
The prompt template is the same as prior works~\cite{24iclr-refusion, 20acl-lmbff}.
The detailed configurations of prefixes are listed in Table~\ref{tab:conf}.
The SemChunk size $c$ is set to 16, the Period size $p$ for ContiguousKV's prefetching is set to 8 layers, and the SubPeriod size $sp$ is set to 4 layers. 
The chunk size used in AS+LRU, AS+H2O+LRU, and IMPRESS is set to 64 tokens.
All experiments use a fixed random seed of 42.
We set the available GPU memory to 10GB and CPU memory to 24GB for all baselines.

\subsection{End-to-End Performance}

\begin{figure*}[t]
\centering
\begin{tabular}{cccc}
  \subfigure[SST-2 (5\%)]{\includegraphics[width=0.25\linewidth]{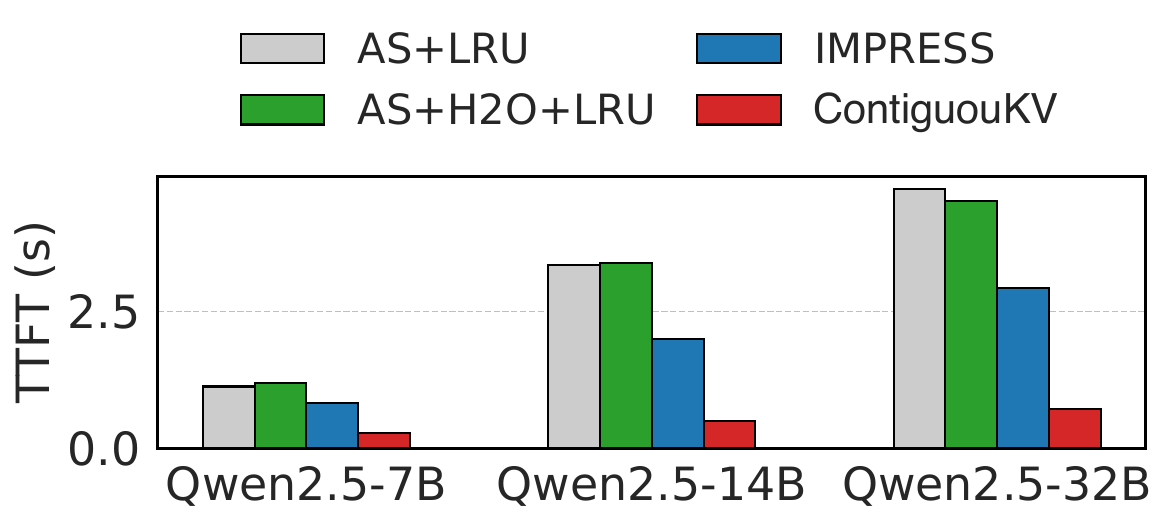}}
  \subfigure[SUBJ (5\%)]{\includegraphics[width=0.25\linewidth]{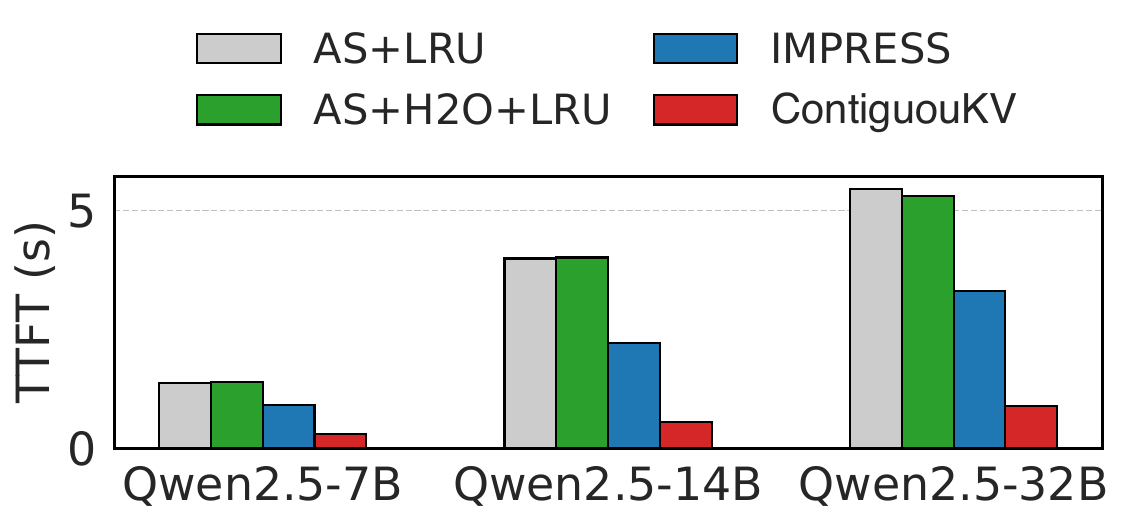}}
  \subfigure[TREC (5\%)]{\includegraphics[width=0.25\linewidth]{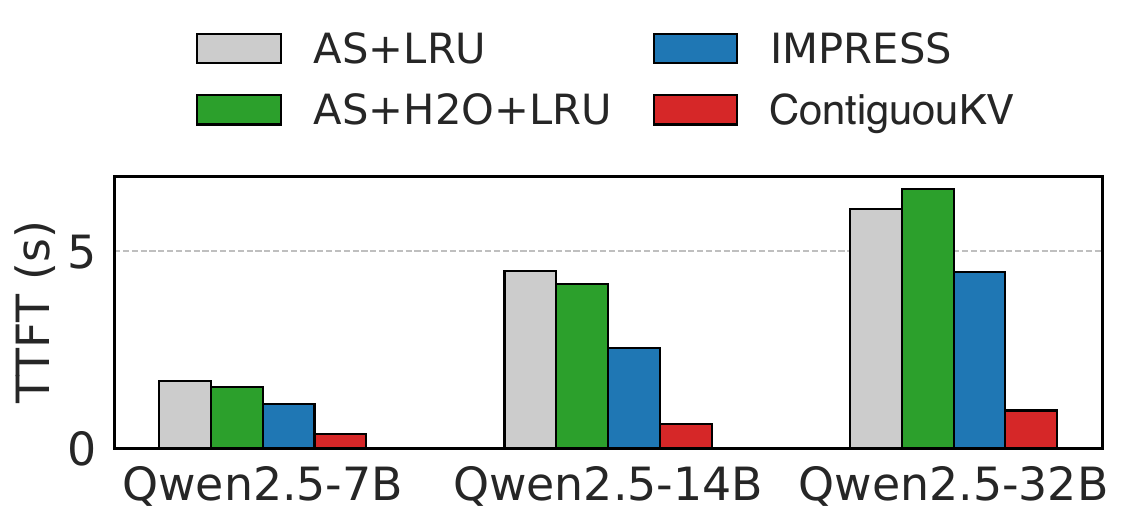}}
  \subfigure[RTE (5\%)]{\includegraphics[width=0.25\linewidth]{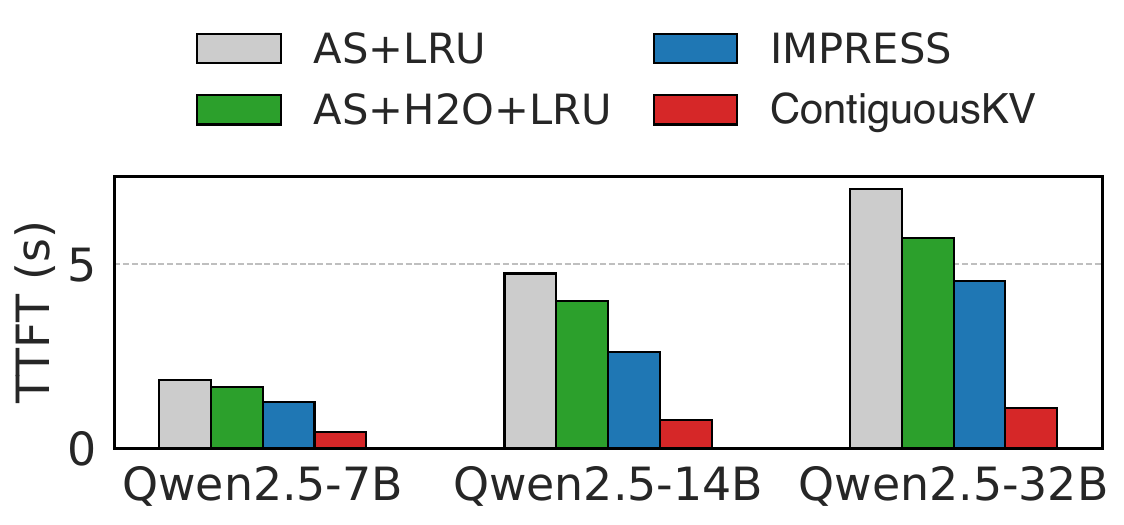}} \\
  \subfigure[SST-2 (25\%)]{\includegraphics[width=0.25\linewidth]{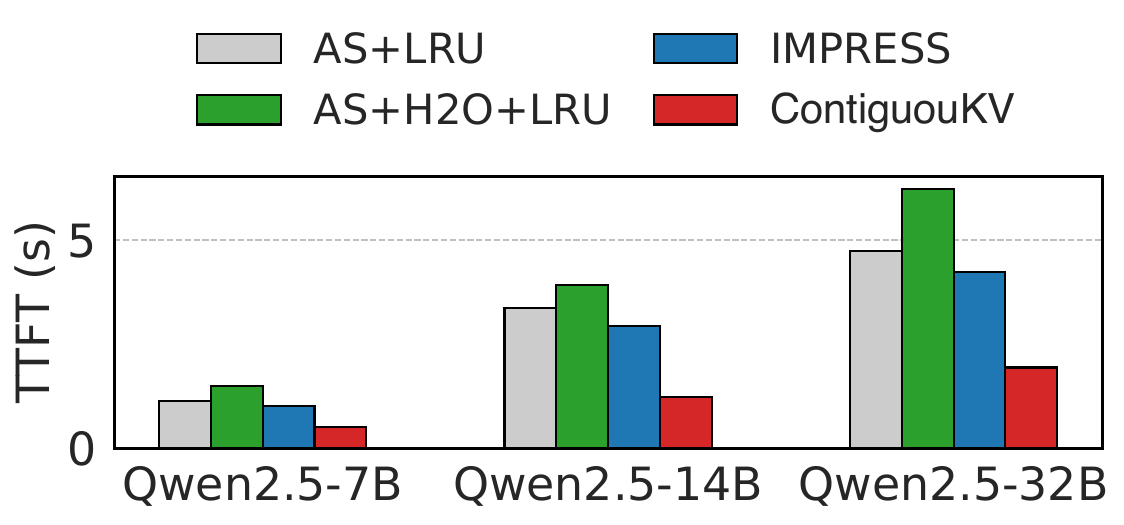}}
  \subfigure[SUBJ (25\%)]{\includegraphics[width=0.25\linewidth]{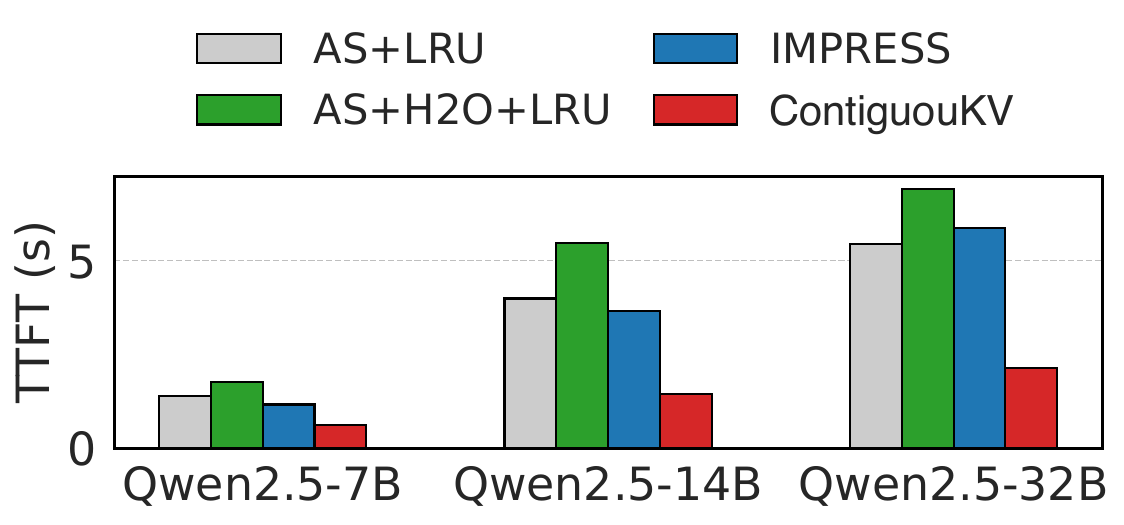}}
  \subfigure[TREC (25\%)]{\includegraphics[width=0.25\linewidth]{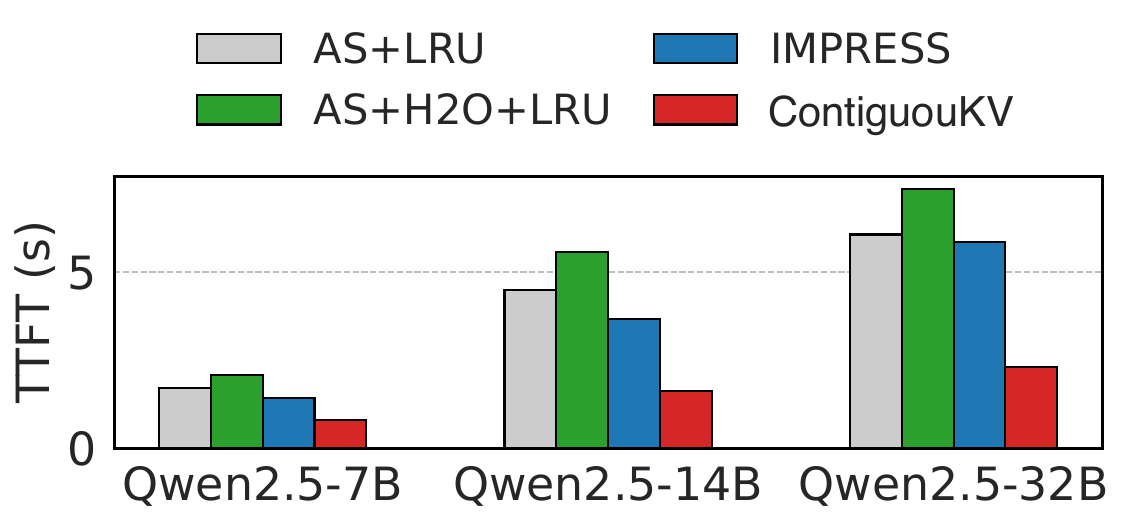}}
  \subfigure[RTE (25\%)]{\includegraphics[width=0.25\linewidth]{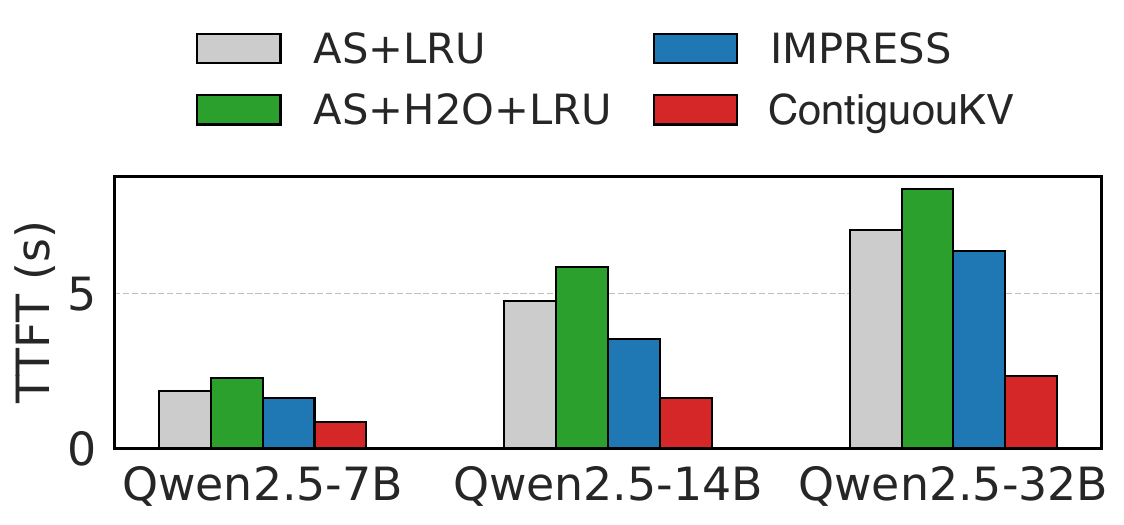}}
\end{tabular}
  \caption{The average TTFT of different inference systems across four datasets and three LLMs. We report results for two KV budget ratios (5\% and 25\%).}
  \label{fig:ttft}
\end{figure*}

\begin{figure}[t]
\centering
\begin{tabular}{cc}
  \subfigure[SST-2]{\includegraphics[width=0.5\linewidth]{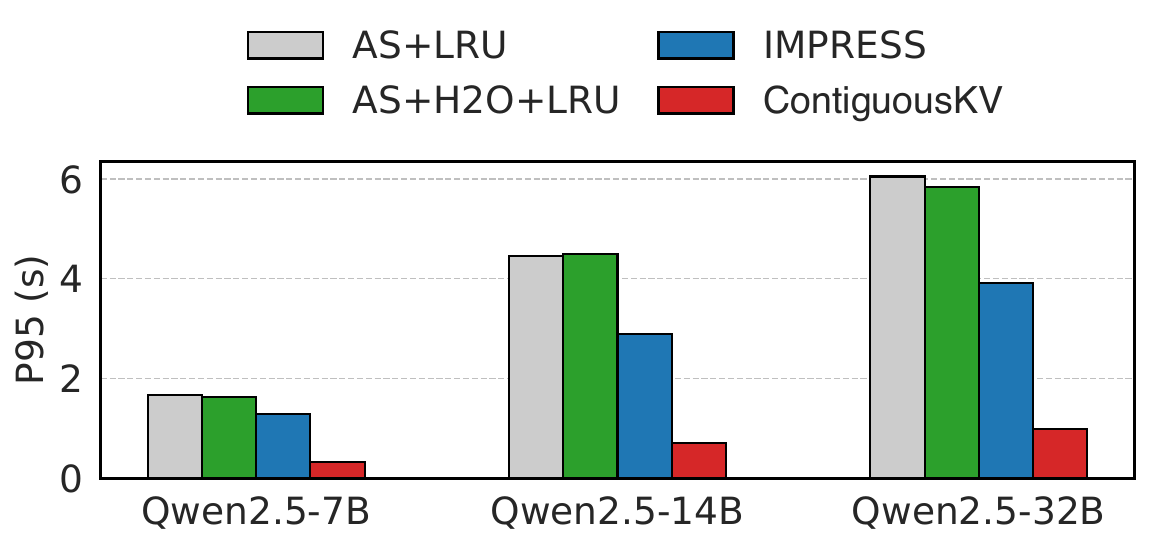}}
  \subfigure[RTE]{\includegraphics[width=0.5\linewidth]{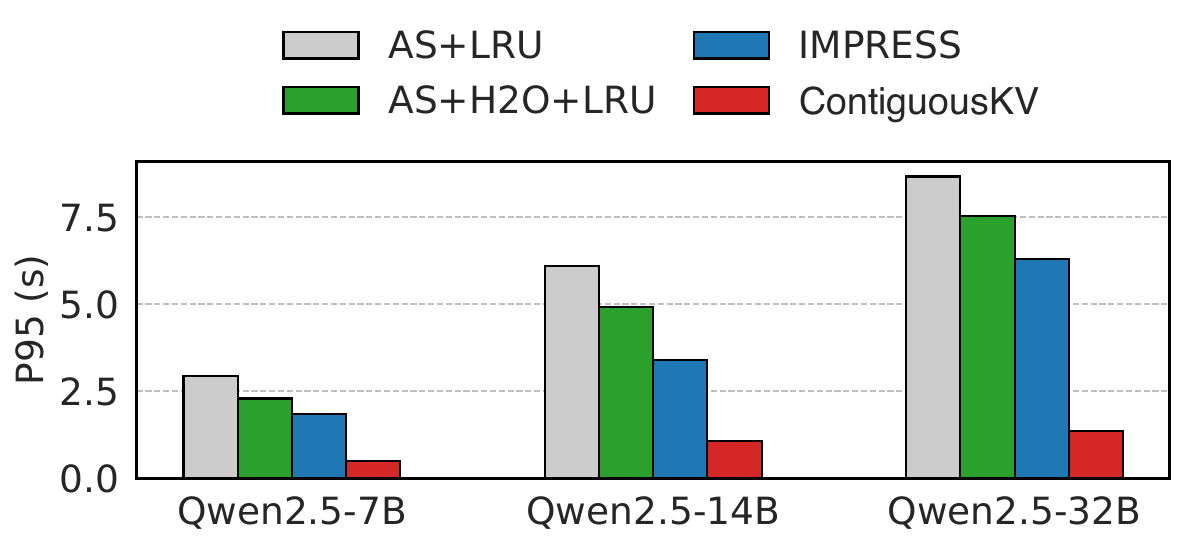}}
\end{tabular}
  \caption{P95 tail latency of different inference systems on representative datasets under 5\% KV cache budget ratio.}
  \label{fig:tail}
\end{figure}

\noindent\textbf{Model Performance.}
Figure~\ref{fig:acc} presents the model performance of our ContiguousKV and existing offloading-based inference systems over four datasets and three LLMs.
Since AS+LRU loads the full KV caches, its model performance doesn't change as the budget ratio increases.
Overall, our ContiguousKV achieves the best performance compared to these systems that adopt the sparse KV cache algorithm among almost all configurations.
Specifically, among different KV budget ratios, ContiguousKV improves the accuracy by 7.69\%, 4.81\%, 3.58\%, and 1.63\% on average compared to IMPRESS, and 10.23\%, 7.08\%, 3.56\%, 2.04\% on average compared to AS+H2O+LFU.
Compared to AS+LRU, ContiguousKV achieves a 5.62\%, 3.56\%, 1.21\%, and 0.04\% decline in accuracy on average across different budget ratios.
When scaling models, ContiguousKV can outperform existing inference systems across all model sizes. 
Especially on larger models (such as Qwen2.5-32B), the improvements of ContiguousKV are more significant, which is 7.71\% and 6.99\% compared to IMPRESS and AS+H2O+LFU, respectively. 
Compared to AS+LRU, ContiguousKV achieves at most 2.94\% performance degradation.
Those results indicate that ContiguousKV's SemChunk can preserve enough semantics for LLMs, thus maintaining acceptable performance.

\noindent\textbf{The Average TTFT and P95 Tail Latency.}
Before the evaluation, we warm up both the GPU and the CPU caches for all baselines.
We evaluate the average TTFT of our ContiguousKV and other systems on two representative budget ratios (5\% and 25\%). 
As shown in Figure~\ref{fig:ttft}, over 5\% KV budget ratio, ContiguousKV achieves 6.16$\times$, 5.83$\times$, and 3.85$\times$ time reductions compared to AS+LRU, AS+H2O+LFU, and IMPRESS, respectively.
A consistent increase in speed at a 25\% budget rate can also be observed in Figure~\ref{fig:ttft}, but these acceleration ratios are less than those on 5\% KV budget ratio.
In terms of model size, ContiguousKV achieves similar acceleration ratios among different model sizes.
These results demonstrate that our ContiguousKV can achieve the maximum speedup at lower budget ratios.
This is because ContiguousKV doesn't introduce read amplifications, thus saving the I/O bandwidth.

We also evaluate the P95 tail latency of our ContiguousKV and other systems on two representative datasets in Figure~\ref{fig:tail}.
To best show the results, we set the KV cache budget ratio to 5\%.
Compared to existing systems, our ContiguousKV significantly reduces the P95 tail latency, 0.42s, 0.89s, and 1.17s on average for different models.
These results show that our ContiguousKV has better stability and robustness.

\subsection{Ablation Studies}
\begin{figure}[t]
\centering
\begin{tabular}{cc}
  \subfigure[SST-2]{\includegraphics[width=0.5\linewidth]{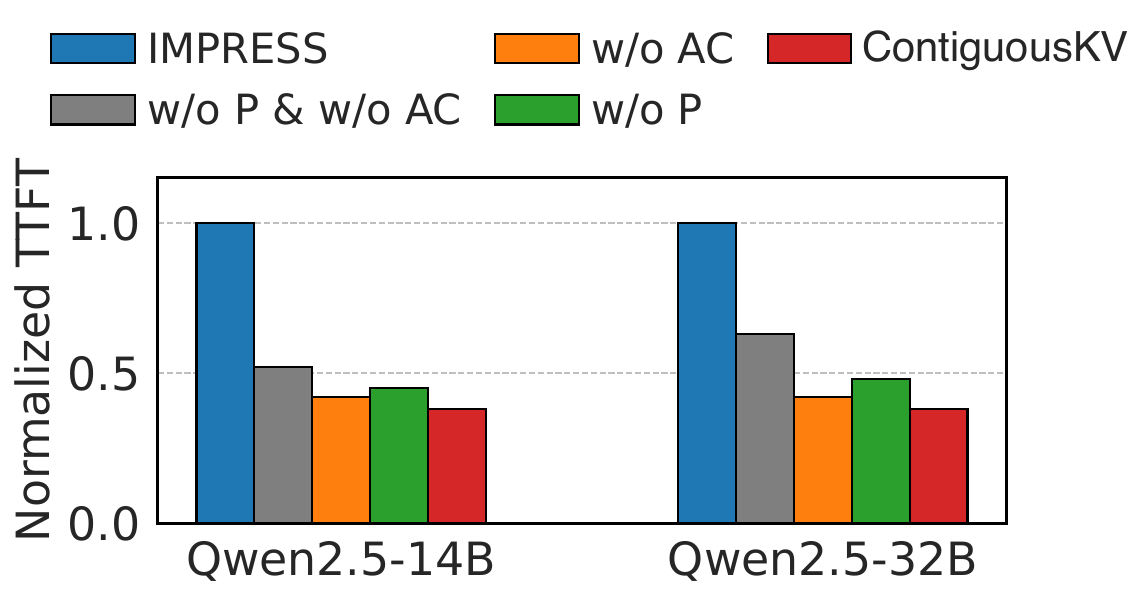}}
  \subfigure[RTE]{\includegraphics[width=0.5\linewidth]{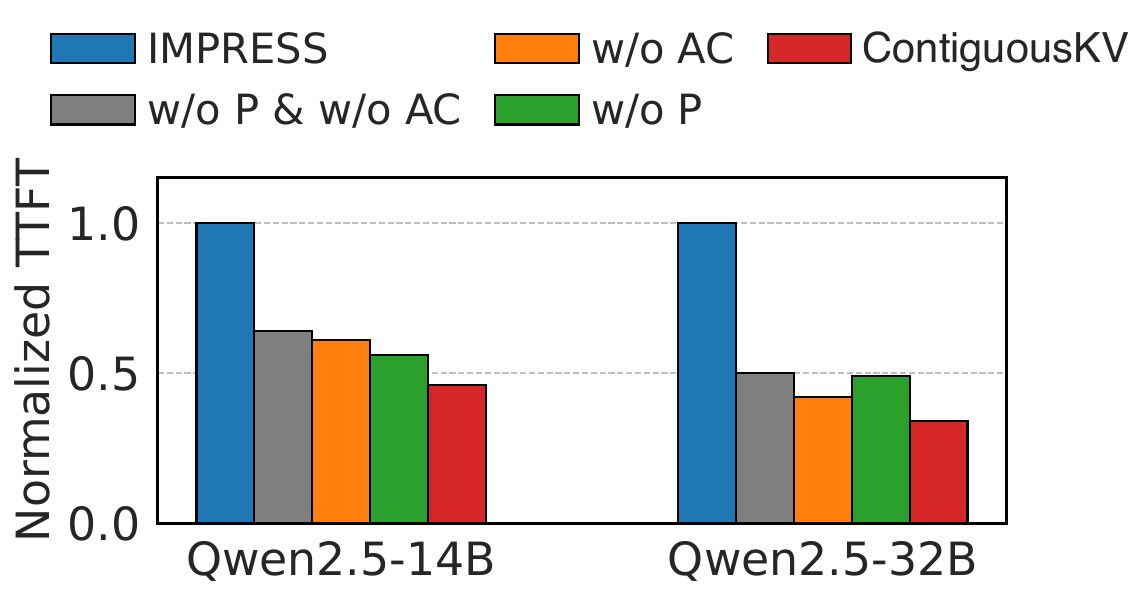}}
\end{tabular}
  \caption{Ablation study of the impact of each optimization.}
  \label{fig:ablation}
\end{figure}

\noindent\textbf{Impact of Individual Optimizations.}
To evaluate the impact of each technique proposed in our ContiguousKV, we conducted an ablation study using two models (Qwen2.5-14B and Qwen2.5-32B) on two datasets (RTE and SST-2).
The budget ratio is 25\%.
Figure~\ref{fig:ablation} shows the ablation study results.
We tested three variants of our ContiguousKV: 
\begin{enumerate}[topsep=0pt,itemsep=-1ex,partopsep=1ex,parsep=1ex]
    \item \textbf{w/o P and w/o AC}: ContiguousKV without Prefetching and using LFU as the default cache policy.
    \item \textbf{w/o AC}: ContiguousKV using LFU as the cache policy.
    \item \textbf{w/o P}: ContiguousKV without Prefetching.
\end{enumerate}
% \textbf{``All''} refers to the default ContiguousKV with all optimizations.

Experimental results in Figure~\ref{fig:ablation} show that, compared to IMPRESS, the major improvement stems from contiguously chunking, which aligns the system granularity to the algorithm granularity.
This is because this optimization can fundamentally mitigate the read amplification.
Both reuse-aware asynchronous prefetching and attention-aware cache management can further accelerate the inference.
However, their contributions vary across models and datasets.
For example, the prefetching contributes more when serving larger models (e.g., Qwen2.5-32B).
This might be due to the fact that larger models require longer computation times, which can overlap more I/O operations.
The new cache policy contributes more to the RTE dataset compared to that on the SST-2 dataset.
This might be related to the difference in the access pattern of each dataset.

\begin{figure}[t]
\centering
  \includegraphics[width=1\linewidth]{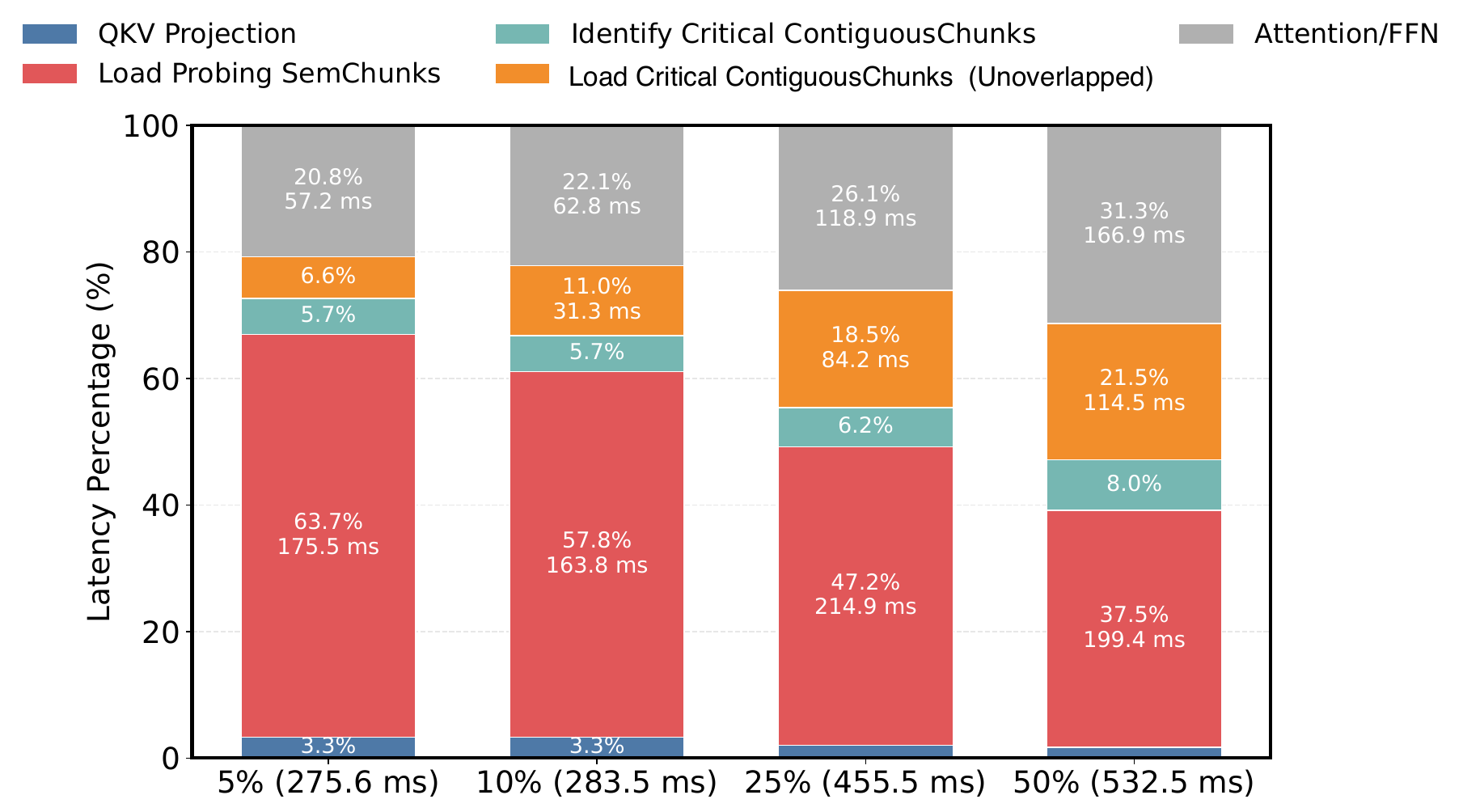}
  \caption{Latency breakdown of  ContiguousKV.}
  \label{fig:breakdown-ours}
\end{figure}

\noindent\textbf{Latency Breakdown.}
Similar to Figure~\ref{fig:breakdown}, we also show the latency breakdown of our ContiguousKV with Qwen2.5-7B on the RTE dataset across different KV cache budget ratios.
We normalize the latency of each stage and report the proportion of each stage to the total latency. 
As shown in Figure~\ref{fig:breakdown-ours}, the proportion of the loading critical SemChunks stage is significantly reduced, resulting in the overall latency reduction.
This is because most of the I/O operations are executed in parallel with computations, which hides the sequential I/O time cost.
Since we don't optimize the loading probing SemChunks stage, which has a similar latency compared to IMPRESS in Figure~\ref{fig:breakdown}, the proportion of this stage increases significantly.
Notably, the computation time cost is also significantly reduced.
This is due to the fact that the alignment eliminates the process of re-assembly sparse tokens from multiple chunks.

% \begin{figure}[t]
% \centering
%   \includegraphics[width=1\linewidth]{figures/io-reduction.pdf}
%   \caption{I/O reductions of our ContiguousKV.}
%   \label{fig:io-reduction}
% \end{figure}

\begin{table}[t]
\caption{Normalized number of tokens loaded from the SSDs in ContiguousKV and IMPRESS. All are normalized based on IMPRESS's results.}
\label{tab:io-reduction}
\begin{center}
\begin{tabular}{lccc}
\toprule
\textbf{Datasets} & \textbf{Models} & \textbf{IMPRESS} & \textbf{ContiguousKV}\\
\midrule
\multirow{2}{*}{\textbf{SST-2}} & 7B & 100\% & 6.21\% \\
 & 14B & 100\% & 6.64\% \\
\multirow{2}{*}{\textbf{RTE}} & 7B & 100\% & 6.00\% \\
 & 14B & 100\% & 5.71\% \\
\bottomrule
\end{tabular}%
\end{center}
\end{table}

\noindent\textbf{I/O Reductions.}
We also conducted an experiment to show the I/O reductions in Table~\ref{tab:io-reduction}.
We quantify this by counting the number of tokens loaded from the SSDs and report the normalized results.
Results indicate that our ContiguousKV can achieve an average reduction of approximately 16.33 times compared to IMPRESS.
This also demonstrates the efficacy of our alignment strategy.

\subsection{Sensitivity Analysis}
% In this section, we conducted several experiments to analyze the sensitivity of hyperparameters (SemChunk size, Period size, and SubPeriod size) used in our ContiguousKV.

\begin{figure}[t]
\centering
\begin{tabular}{cc}
  \subfigure[Qwen2.5-7B]{\includegraphics[width=0.5\linewidth]{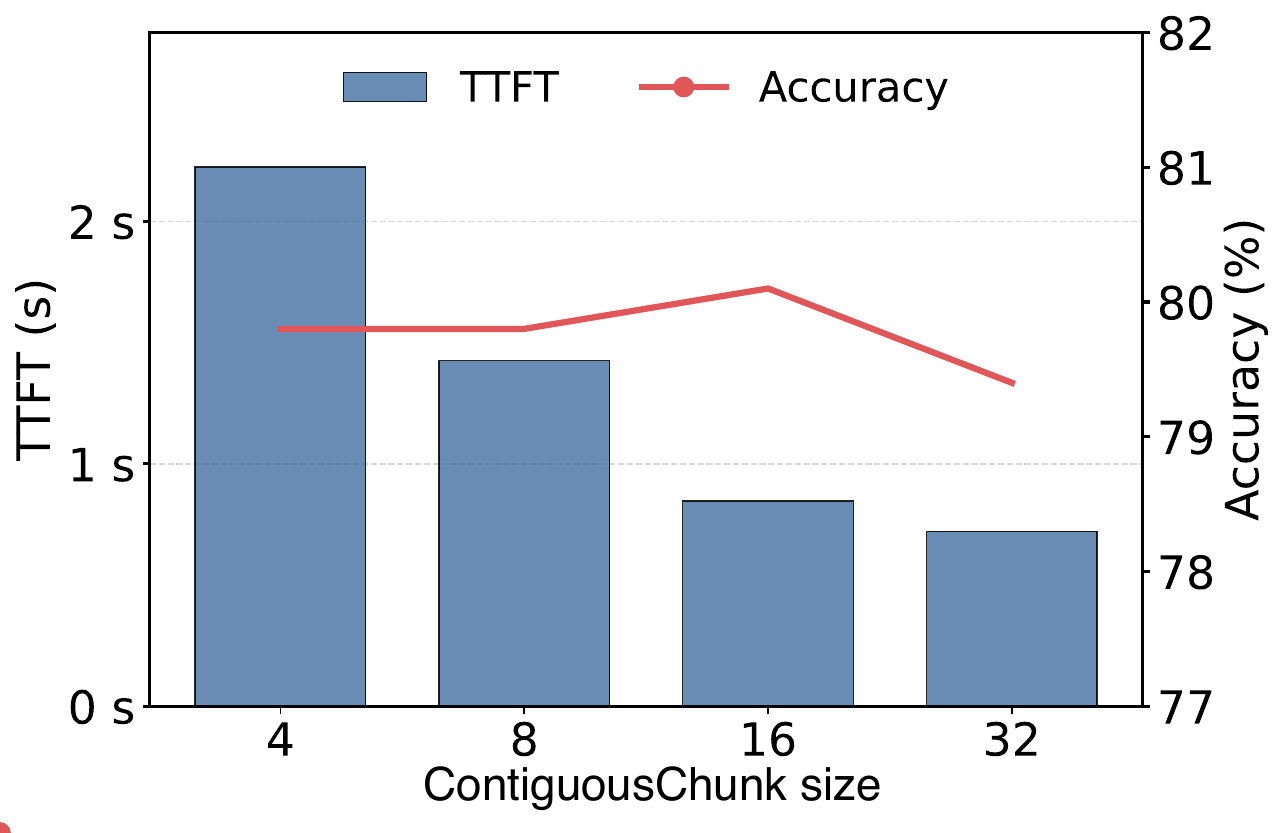}}
  \subfigure[Qwen2.5-14B]{\includegraphics[width=0.5\linewidth]{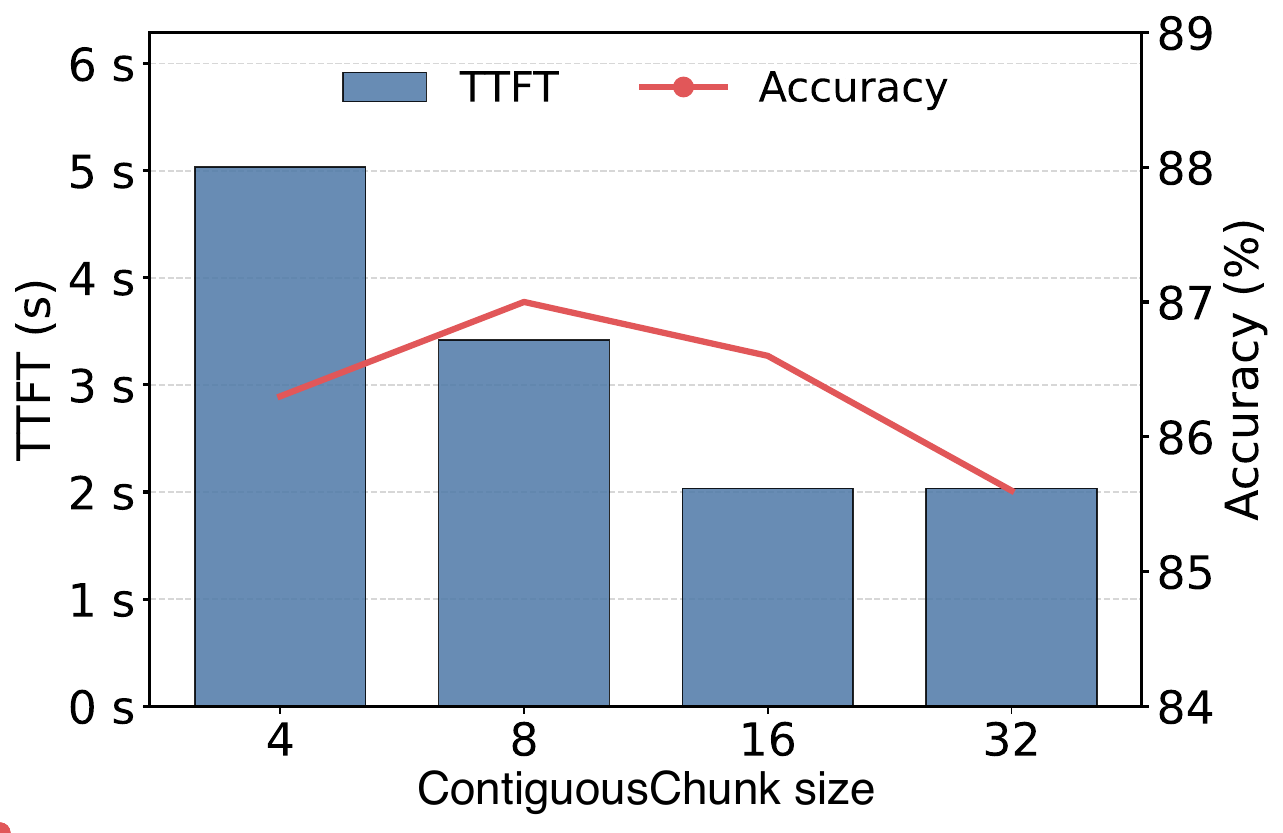}}
\end{tabular}
  \caption{Trade-off between accuracy and TTFT under different SemChunk sizes.}
  \label{fig:semchunk}
\end{figure}

\noindent\textbf{SemChunk Size.}
Due to our alignment between system granularity and algorithm granularity, the SemChunk size not only impacts the model performance but also the model efficiency. 
We evaluate the accuracy and the average TTFT of ContiguousKV with different SemChunk sizes, ranging from 4 to 32.
As shown in Figure~\ref{fig:semchunk}, as the SemChunk size increases, the model performance first exhibits an upward trend, then declines.
This is because rich semantics can help LLMs' understanding capability, but too much information may introduce noise.
Regarding efficiency, since we have aligned system granularity to algorithm granularity, a larger SemChunk size can utilize more I/O bandwidth, thereby accelerating inference.
In this paper, we choose 16 as the SemChunk size, as it makes a good trade-off between accuracy and efficiency.
% Notably, for Qwen2.5-14B, the TTFT using a SemChunk size of 32 is similar to that of a SemChunk size of 16.
% This might be due to the fact that the SemChunk size has reached the maximal I/O bandwidth.

\begin{figure}[t]
\centering
\begin{tabular}{cc}
  \subfigure[Qwen2.5-7B]{\includegraphics[width=0.5\linewidth]{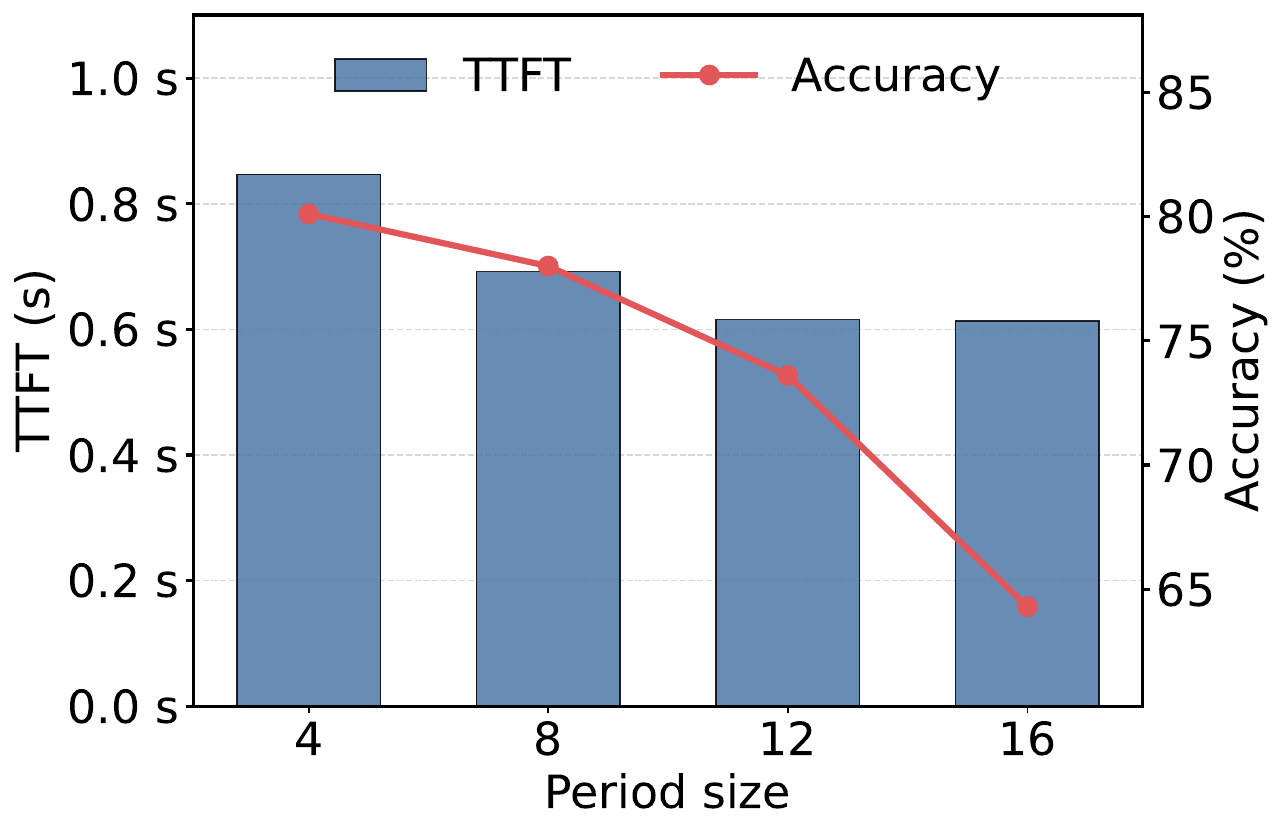}}
  \subfigure[Qwen2.5-14B]{\includegraphics[width=0.5\linewidth]{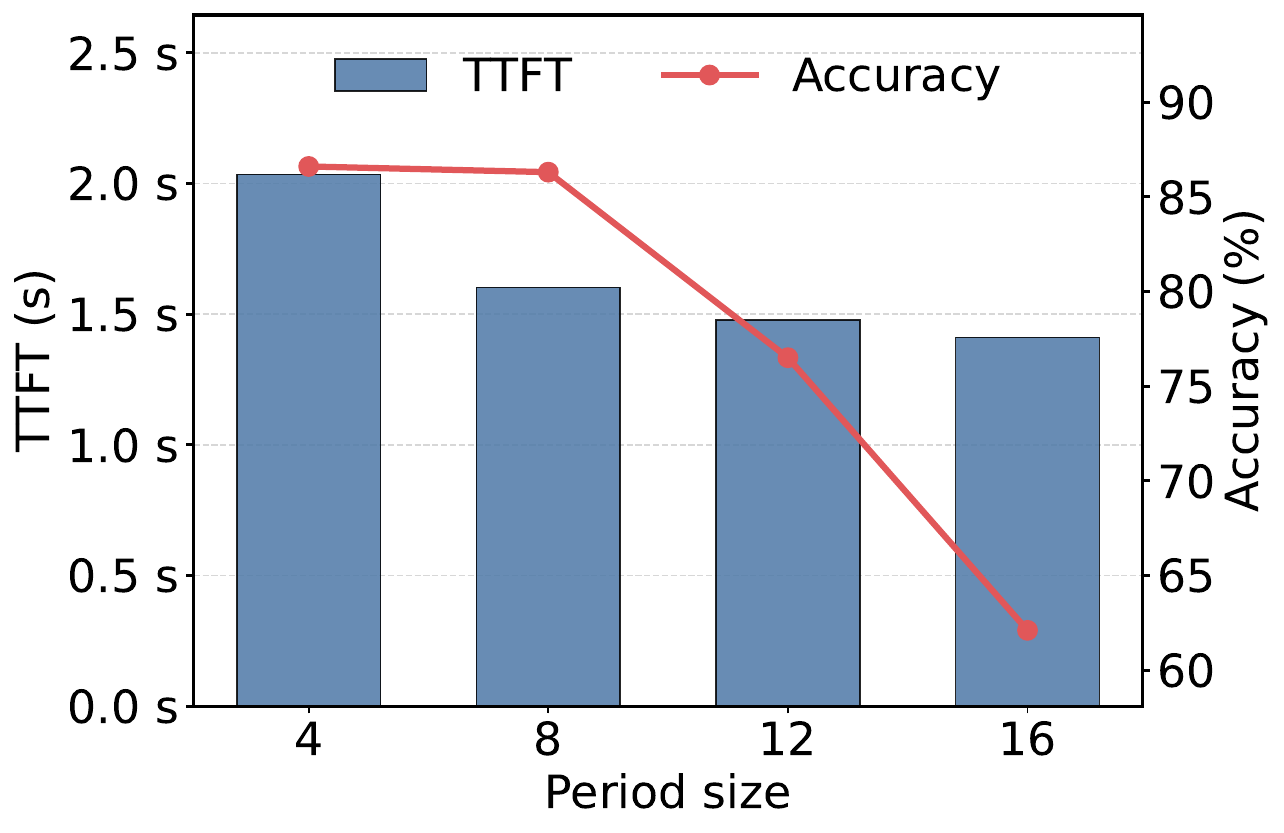}}
\end{tabular}
  \caption{Trade-off between accuracy and TTFT under different Period sizes.}
  \label{fig:period}
\end{figure}

\noindent\textbf{Period Size.}
Period size determines the number of layers where the critical SemChunk indices are shared.
With our prefetching mechanism, a larger Period size can overlap more I/O operations with computations.
We also evaluate the accuracy and the average TTFT of ContiguousKV with different Period sizes, ranging from 4 to 16.
As shown in Figure~\ref{fig:period}, as the Period size increases, the model performance degrades correspondingly, while the model efficiency is enhanced.
This is because the reuse strategy is an approximate process that can accumulate losses as the number of reuse layers increases.
Therefore, in this paper, we choose a Period size of 8, also for a better trade-off.

\begin{figure}[t]
\centering
\begin{tabular}{cc}
  \subfigure[Prefix Size]{\includegraphics[width=0.5\linewidth]{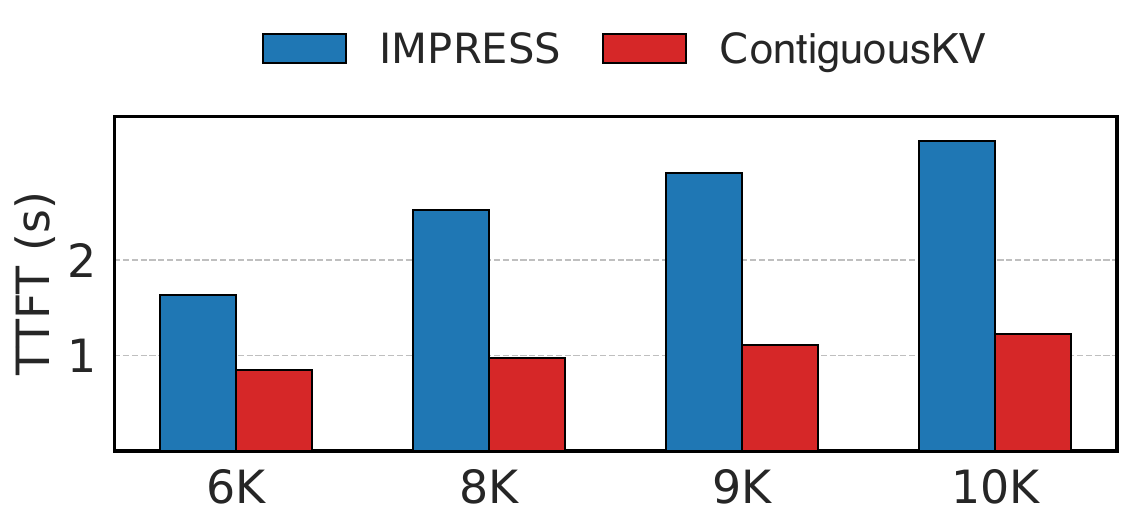}}
  \subfigure[SubPeriod Size]{\includegraphics[width=0.5\linewidth]{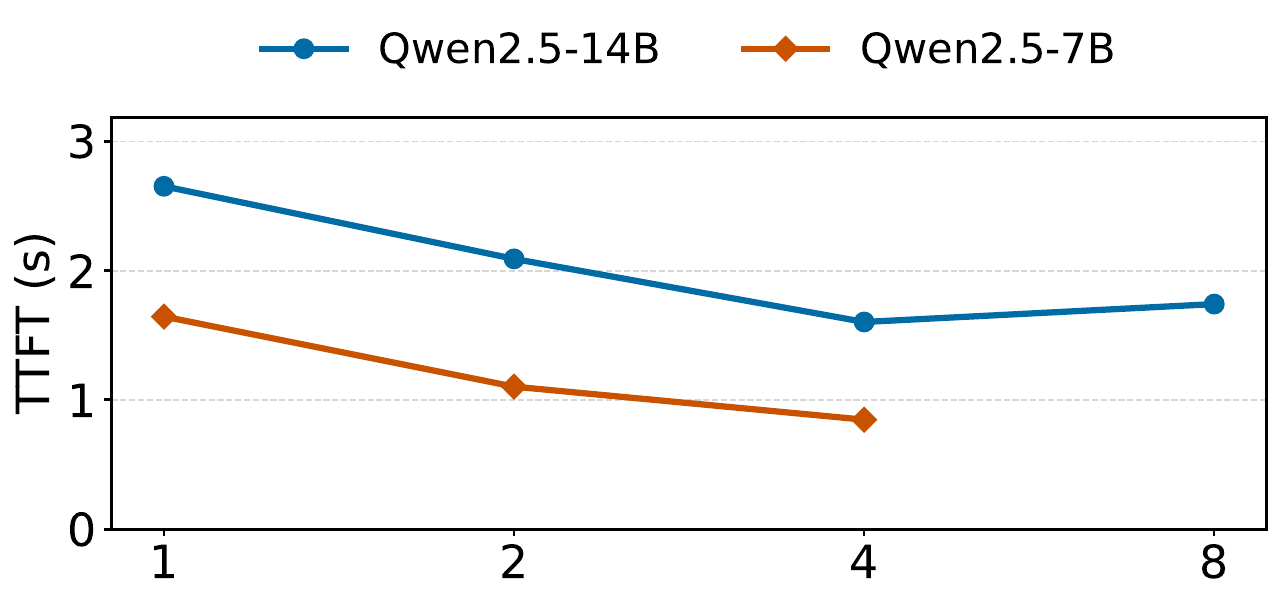}}
\end{tabular}
  \caption{The average TTFT under different SubPeriod sizes and different prefix sizes.}
  \label{fig:subperiod}
\end{figure}

% \noindent\textbf{SubPeriod Size.}
% SubPeriod size, as a hyperparameter, determines the degree of overlap between I/Os and computations.
% We evaluate the average TTFT under different SubPeriod sizes in Figure~\ref{fig:subperiod}.
% Since Qwen2.5-7B only has 28 layers, we vary the SubPeriod size from 1 to 4.
% Results show that as the SubPeriod size increases, the TTFT is reduced.
% However, for Qwen2.5-14B, the TTFT increases at the SubPeriod size of 8.
% This is because in this experiment, we set the Period size to 8.
% When the SubPeriod size is the same as the Period size, there is no overlap within a Period.
% Notably, although a small SubPeriod size can perform as much prefetching as possible, its computation time is relatively small, so that it is unable to overlap the I/Os.
% Therefore, in this paper, we choose a SubPeriod size of 4 to maximally overlap I/Os and computations.

\noindent\textbf{Prefix Size Scalability.} 
We further test the scalability of our ContiguousKV and IMPRESS by varying the shared prefix length under a 25\% KV cache budget with Qwen2.5-7B. 
Results in Figure~\ref{fig:subperiod}(a) show that ContiguousKV consistently outperforms IMPRESS, and its performance advantage grows with longer prefixes. 
For instance, with a 10K-token prefix, ContiguousKV achieves a 2.6× speedup. 
This demonstrates that our system efficiently handles increasing context lengths without the severe latency inflation seen in prior systems.

\noindent\textbf{SubPeriod Size.} 
We evaluate the average TTFT of our ContiguousKV under varying Subperiod sizes with two models. 
As shown in Figure~\ref{fig:subperiod}(b), a larger SubPeriod size generally reduces TTFT by enabling better overlap between I/O and computation. 
We choose a size of 4 in this paper, as it optimally balances prefetching aggressiveness with available compute time for overlap.
The SubPeriod size equal to the full Period size (e.g., 8) would degrade efficiency as it eliminates intra-period pipelining.

\section{Related Work}
\label{sec:related}

\noindent\textbf{Offloading-based Inference Systems.} 
To accommodate the massive KV caches during long-context LLM inference, a line of research explores offloading strategies. 
One category of work~\cite{25sigmod-pqcache, 25acl-f-a2ats} extends the KV cache from GPU to CPU main memory, treating it as a larger secondary cache tier. 
These works further organize these CPU-resident caches into retrieval databases to enable sparse attention. 
While this alleviates GPU memory pressure, its capacity remains bounded by DRAM size, and the retrieval index often needs reconstruction for new input patterns, introducing a considerable overhead. 
Another category of systems~\cite{25fast-impress, 24atc-as, 24osdi-infinigen, 25asplos-vattention}, more closely aligned with our work, employs a three-tier storage hierarchy (GPU memory, CPU memory, and SSD) to provide near-unlimited capacity. 
However, these systems manage data in coarse, fixed-size blocks for I/O amortization, which leads to severe read amplification when combined with fine-grained, semantic-aware pruning algorithms. 
ContiguousKV directly addresses this fundamental granularity mismatch by introducing the ContiguousChunk as a unified unit for pruning, storage, transfer, and caching, thereby eliminating read amplification and its associated I/O waste.

\noindent\textbf{KV Cache Pruning and Quantization.}
KV cache pruning methods identify and retain only a subset of important KV pairs. 
Token-level pruning techniques, such as H2O~\cite{23nips-h2o}, StreamingLLM~\cite{24iclr-streamingllm}, and SnapKV~\cite{24nips-snapkv}, score and select individual tokens. 
While effective in reducing size, operating at token granularity can disrupt local semantic coherence, potentially leading to a noticeable degradation in generation quality. 
Recent chunk-level methods like ChunkAttention~\cite{24acl-chunkattention} and ChunkKV~\cite{25nips-chunkkv} group consecutive tokens and perform importance selection at the chunk level, better preserving contextual semantics, an insight ContiguousKV adopts and extends into system design. 
Orthogonally, quantization techniques~\cite{24icml-kv-quant-1, 24acl-kv-quant-2, 25acl-kv-quant-3, 24nips-kv-quant-4} reduce the bit-width of each KV element (e.g., from FP16 to 8-bit or 4-bit) to shrink the memory footprint. 
However, the absolute savings diminish as context length grows into the tens of thousands of tokens, and quantization alone cannot solve the fundamental I/O bottleneck when the entire quantized cache still must be loaded from slow storage. 
ContiguousKV's pruning-at-ingest and selective loading fundamentally reduce the amount of data that needs to be transferred, a benefit complementary to quantization.

\noindent\textbf{Other LLM Serving Systems.}
A broad spectrum of work optimizes other aspects of LLM serving. Systems like vLLM~\cite{23sosp-vllm} and SGLang~\cite{24nips-sglang} focus on efficient GPU memory management for the dynamic KV cache during decoding, primarily through paging and advanced scheduling. Other projects, such as DistServe~\cite{24osdi-distserve} and Mooncake~\cite{25fast-mooncake}, propose disaggregating the prefill and decoding phases across a cluster to improve aggregate goodput. 
These optimizations are largely orthogonal to ContiguousKV’s focus on minimizing I/O latency during the Re-Prefill phase for workloads with shared, offloaded prefix KV caches. 
ContiguousKV’s techniques can be integrated with these systems to further improve end-to-end performance.

\section{Conclusion}
\label{sec:conclusion}
In this paper, we identified and addressed the critical Re-Prefill bottleneck in LLM serving systems that handle shared-context workloads. Through detailed analysis, we uncovered two fundamental inefficiencies: a granularity mismatch between semantic pruning and physical I/O operations that causes read amplification, and sequential dependencies in the Re-Prefill process that create resource underutilization. Our system, ContiguousKV, introduces the ContiguousChunk abstraction to align I/O granularity with algorithmic granularity, eliminating wasteful data transfers by ensuring that each I/O operation fetches precisely the semantically cohesive units required. We developed a two-level prefetching mechanism that exploits both intra-period and inter-period similarities in attention patterns across transformer layers, enabling asynchronous, speculative loading of KV cache data while the current layer computes. This pipelining breaks the sequential dependency that previously created idle bubbles. Additionally, our attention-aware cache management policy dynamically prioritizes chunks based on the model's own attention scores, ensuring that the most semantically valuable prefix data remains in faster memory tiers. Evaluation with Qwen2.5 models demonstrates that ContiguousKV reduces TTFT by 3.85$\times$ compared to state-of-the-art solutions.

% %-------------------------------------------------------------------------------
% \section*{Acknowledgments}
% %-------------------------------------------------------------------------------

% The USENIX latex style is old and very tired, which is why
% there's no \textbackslash{}acks command for you to use when
% acknowledging. Sorry.

% %-------------------------------------------------------------------------------
% \section*{Availability}
% %-------------------------------------------------------------------------------

% USENIX program committees give extra points to submissions that are
% backed by artifacts that are publicly available. If you made your code
% or data available, it's worth mentioning this fact in a dedicated
% section.

%-------------------------------------------------------------------------------
% \newpage
\bibliographystyle{plain}
\bibliography{ref}

%%%%%%%%%%%%%%%%%%%%%%%%%%%%%%%%%%%%%%%%%%%%%%%%%%%%%%%%%%%%%%%%%%%%%%%%%%%%%%%%
\end{document}